\documentclass[12pt]{article}
\pdfoutput=1
\usepackage{jcapmod}
\usepackage{booktabs}

\usepackage[english]{babel}
\usepackage{amsmath,amssymb,amsbsy,amstext, amsthm,xcolor}
\usepackage{graphicx}
\usepackage{amsfonts}
\usepackage{amssymb}
\usepackage{float}
\usepackage[utf8]{inputenc}
\RequirePackage{color}

\usepackage{colortbl}
\definecolor{green2}{cmyk}{0, 1, 0.5, 0}
\definecolor{lightgreen}{cmyk}{0.2, 0, 0.2, 0.2}
\definecolor{lightgray}{cmyk}{0.1,0.2,0,0.1}
\definecolor{lightgray2}{cmyk}{0.4,0.4,0,0.8}
\definecolor{black}{cmyk}{1.0,1.0,1.0,1.0}

\allowdisplaybreaks[1]

\usepackage{colortbl}
\definecolor{lightgreen}{cmyk}{0.2, 0, 0.2, 0.2}
\definecolor{lightgray}{cmyk}{0.1,0.2,0,0.1}
\definecolor{lightgray2}{cmyk}{0.1,0.1,0,0.1}

\makeatletter
\newlength{\apb@width}
\newcommand{\autoparbox}[2][c]{\settowidth{\apb@width}{#2}\parbox[#1]{\apb@width}{#2}}

\makeatother

\numberwithin{equation}{section}

\def\beq{\begin{equation}}
\def\eeq{\end{equation}}

\def\bea{\begin{eqnarray}}
\def\eea{\end{eqnarray}}
\def\eg{{\it e.g.~}}

\def\ie{{\it i.e.~}}
\def\d{{\rm d}}

\def\d{{\rm d}}

\def\nn{\nonumber}

\def\sgm{\sigma}

\def\Mp{M_{\rm pl}}

\def\fr{\frac}

\def\0{{\boldsymbol 0}}

\def\fr{\frac}

\usepackage{setspace} 
\begin{document}

\begin{titlepage}

\setcounter{page}{1} \baselineskip=15.5pt \thispagestyle{empty}

\bigskip\

\vspace{1cm}
\begin{center}

{\fontsize{20}{28}\selectfont  \sffamily \bfseries {Mechanisms for Primordial Black Hole Production in String Theory  \\ }}

\end{center}

\vspace{0.2cm}

\begin{center}
{\fontsize{13}{30}\selectfont Ogan \"Ozsoy$^{\clubsuit}$, Susha Parameswaran$^{\spadesuit}$, Gianmassimo Tasinato$^{\clubsuit}$, Ivonne Zavala$^{\clubsuit}$}
\end{center}

\begin{center}

\vskip 8pt
\textsl{$^\clubsuit$ College of Science, Swansea University, Swansea, SA2 8PP, UK}\\
\textsl{$^\spadesuit$ Department of Mathematical Sciences, University of Liverpool, Liverpool, L69 7ZL, UK}
\vskip 7pt

\end{center}

\vspace{1.2cm}
\hrule \vspace{0.3cm}
\noindent {\sffamily \bfseries Abstract} \\[0.1cm]
We consider  mechanisms for producing a significant population of primordial black holes (PBHs) within string inspired single field models of inflation.  
 The production of PBHs requires a large amplification in the power spectrum of curvature perturbations between scales associated with CMB and PBH formation.   
In principle, this can be achieved by temporarily breaking the slow-roll conditions during inflation. In this work,
we identify two string  setups that can realise this process.  In string axion models of inflation, subleading non-perturbative effects can superimpose steep cliffs and gentle plateaus onto the leading axion potential.  The cliffs can momentarily   violate the slow-roll conditions, and the plateaus can lead to phases of ultra slow-roll inflation.  We thus achieve a string motivated model which both matches the Planck observations at CMB scales
 and produces a population of light PBHs, which can account for 
    an order one fraction of  dark matter.  In DBI models of inflation, a sharp increase in the speed of sound sourced by a steep downward step in the warp factor can drive the amplification.  In this scenario, discovery of PBHs could indicate non-trivial dynamics in the bulk, such as flux-antibrane annihilation at the tip of a warped throat.
\vskip 10pt
\hrule

\vspace{0.6cm}
 \end{titlepage}

 \tableofcontents
 
\newpage
\section{Introduction}
In  the early 1970s, Stephen Hawking proposed that overdense inhomogeneities in the very early Universe could gravitationally collapse to form Primordial Black Holes (PBHs) \cite{Hawking:1971ei,Carr:1974nx}.  Assuming the Planck length as the minimal possible Schwarzchild radius, gives a lower limit for black holes masses of about $10^{-5}$g.  Moreover, this mechanism  -- in contrast to the astrophysical process from dying stars --  allows for populations of black holes spanning a vast range of masses, from the Planck scale to supermassive and beyond.  Constraints on PBHs abundances continue to improve, but leave viable windows, especially when astrophysical uncertainties are taken into account (see \cite{Carr:2016drx,Sasaki:2018dmp} for recent reviews).  

  PBHs could provide a significant fraction -- or indeed all -- of the mysterious Dark Matter that dominates cosmic structures in the present day Universe.  Stimulated both by the absence of signatures for well-motivated particle candidates for Dark Matter and by the first detection of gravity waves from merging black holes, this idea has recently taken new flight.  The principal modern process for PBH production is the collapse of curvature perturbations generated during early Universe cosmic inflation, when the relevant scales re-enter the horizon during the radiation era \cite{GarciaBellido:1996qt}.  However, the big challenge here is that the amplitude of these inflationary curvature perturbations at the large scales probed by the CMB has been constrained by observations to be tiny, $A_s \sim 10^{-9}$ at the pivot scale $k=0.05$ Mpc$^{-1}$.  In order for inflationary perturbations to lead to a significant PBH population there must be some mechanism at work that enhances the power spectrum to around $A_s \sim 10^{-2}$ at some smaller scale.  For example, a large enhancement at scales around $k \sim 10^{12} - 10^{14}$ Mpc$^{-1}$ would lead to PBHs masses in the window $10^{-17} M_{\odot} \lesssim M_{PBH}  \lesssim 10^{-13} M_{\odot}$, which could be a dominant component of Dark Matter consistently with current observational constraints\footnote{A possible obstacle for PBH's as Dark Matter in this mass range is the constraints arise  due to capture of PBH by stars during their formation which may limit the PBH abundance to no more than one percent of the Dark Matter density \cite{Capela:2014ita}.}.



Many recent works have explored how such an enhancement could be achieved.  As current CMB observations are consistent with the simplest, single field slow-roll models, much of the focus has been to explore scenarios producing PBHs in single field inflation \cite{Garcia-Bellido:2017mdw,Ezquiaga:2017fvi,Ballesteros:2017fsr,Hertzberg:2017dkh,Kannike:2017bxn}.  
A possibility is that
 the inflaton rolls down a slow-roll plateau, probed by the CMB observations, followed by a near-inflection point which leads to a phase of ultra slow-roll.  Roughly, the extreme flattening of the inflationary potential around the inflection point leads to an enhanced amplitude of scalar perturbations.   
Whilst such a potential can certainly be manufactured within effective field theory, ultimately its shape -- and robustness against quantum corrections -- would have to be explained within a fundamental theory, like string theory.  For example, an attempt towards embedding this model within string theory via fine-tuned string-loop corrections in fibre inflation has been made very recently in \cite{michele}, although it has not yet proved possible to fine-tune a setup such that cosmological perturbations both match the CMB and produce significant PBH Dark Matter.

The purpose of this paper is to identify well-motivated string mechanisms  that can lead to an amplification of curvature perturbations and thus significant PBH production in the early Universe.  To do so, it is helpful to recall  the physical conditions  underlying  this amplification, first discussed by Leach et al. \cite{Leach:2000yw,Leach:2001zf}\footnote{Note however, the first toy model that can amplify the power in scalar fluctuations is proposed by Starobinsky  \cite{Starobinsky:1992ts}.}.  In slow-roll inflation, the equation that governs the curvature perturbation takes the form of a damped harmonic oscillator.  Solutions for super-horizon modes correspond to a linear combination of a constant and a decaying mode.  However, if the slow-roll conditions are broken, then the friction term in the damped harmonic oscillator equation can become a driving term, so that the decaying mode becomes a growing mode, and enhances the curvature perturbation on  scales beyond the horizon.  This enhancement continues until the slow-roll conditions are restored. (A similar mechanism has been used to build single field models with large squeezed non-Gaussianity \cite{Chen:2013aj}.)  

The first  possible scenario that we propose is based upon an earlier observation on how subleading, non-perturbative effects can affect axion inflation \cite{Parameswaran:2016qqq} \footnote{Other
papers studying PBH production in axion inflation are \cite{Linde:2012bt,Bugaev:2013fya,Garcia-Bellido:2016dkw,Domcke:2016bkh,Cheng:2018yyr}, 
although they do not work in a single field system, since they  couple the axion to gauge fields or include a spectator sector coupled to gauge fields.}.  In string axion inflation, the perturbative axion shift symmetry is broken spontaneously by background vevs (\eg fluxes), such as in axion monodromy \cite{SW,McAllister:2008hb} and/or non-perturbative effects (\eg string instantons), leading to large field inflation models with monomial or cosine (``natural inflation'' \cite{Freese:1990rb}) potentials.  In \cite{Parameswaran:2016qqq}, we noted that subleading non-perturbative effects -- if sufficiently large -- can superimpose periodically steep cliffs and gentle plateaus onto the underlying potential.  The overall effect shown there was to restore axion inflation into the favour of current CMB observations, by allowing sufficient efolds of inflation to be obtained with smaller field ranges and thus lowering the tensor-to-scalar ratio, even achieving natural inflation with sub-Planckian axion decay constants.  Here we note that such an inflationary potential can also quite naturally give rise to an enhancement in the scalar power spectrum at small scales, as the inflaton meets successively shallower plateaus on its roll towards the global minimum.  Depending on the parameters, the cliffs may temporarily halt inflation, and the plateaus may include (near-)inflection points around which ultra slow-roll inflation occurs.  By connecting to the heuristic arguments above, and analysing numerically the cosmological perturbations, we show that such models can both lie within  2$\sigma$ constraints from Planck results, and produce a population of light PBH (about $10^{-16}-10^{-15} M_\odot $)  that provides a signification fraction of Dark Matter.

The second string motivated  mechanism for the amplification of power during inflation that we propose is based on  single field models of inflation with non-canonical kinetic terms, as for example DBI inflation \cite{ST,AST}.  To understand what features are required within these models, we extend the Leach et al. argument on amplification of super-horizon modes to include varying speed of sound.  We find that a driving term emerges in the mode equation if the speed of sound increases sufficiently fast.  A simple way to achieve this is if the warp factor experienced by the D-brane sourcing inflation features a sufficiently large\footnote{
See  \cite{Bean:2008na,MHA} for an analyisis of the  effects on brane inflation of the tiny, sharp steps in the warp factor caused by quantum corrections in Seiberg duality cascades \cite{HT}.}  and steep  step during the inflationary trajectory.  A large step in the warp factor could occur for instance, if the D-brane travels down a throat within a throat, sourced by two separated stacks of D-branes \cite{Franco:2005fd, Cascales:2005rj}.  

 Another scenario could be that during the D-brane's journey down the throat, an instability towards brane-antibrane or flux-antibrane \cite{Kachru:2002gs} annihilation occurs at the tip of the throat.  In this scenario, observations of a population of PBHs would reveal interesting properties of branes exploring the geometry of the string compactification.

The paper is organised as follows.  In the following section, we provide a brief review of the basic physical
 mechanism which leads to the enhancement of the amplitude of curvature perturbation during inflation, subsequently leading to the production of PBHs.
In Section 3, we review and develop models  of string axion inflation including subleading non-perturbative effects, describing their background evolution and predictions at the CMB scales.  We then investigate numerically the amplification of scalar power spectrum at small scales.  In particular, we provide two explicit models which both concord with CMB observations and produce PBHs sufficient to explain a significant fraction of Dark Matter.  In Section 4, we turn our attention to DBI models of inflation with a large step feature in the warp factor.  We study the background evolution and the equation governing the curvature perturbations, and thus provide heuristic arguments for the amplification of curvature perturbations, postponing a detailed numerical analysis of the perturbations to future work.  Finally, in Section 5 we present our conclusions and point towards several open questions to which our work leads.

We will use natural units, $\hbar=c=1$, with reduced Planck mass $\Mp^2 =(8\pi G)^{-1}$. Our metric signature is mostly plus  $(-,+,+,+)$. The background metric is a FRW universe with line element $d s^2\,=\,
-d t^2+a^2(t)\,d \vec x^2\,=\, a^2(\tau)\,\left( 
- d \tau^2+d \vec x^2
\right)
$.
The overdots and primes on time dependent quantities  denote derivatives with respect to coordinate time $t$ and conformal time $\tau$, respectively. During inflation, we take $a(\tau)= 1/(-H\tau)^{1+\epsilon}$ with $H$ is the physical Hubble rate.

\section{Enhancing the amplitude of curvature fluctuations in single field inflation}
\label{SecHeu}

As  discussed in the introduction, the production of PBHs from inflation requires that
 the spectrum of curvature fluctuations has to increase by a factor of $\sim 10^{7}$ in its amplitude at scales well below CMB  scales. 
Such enhanced primordial curvature fluctuations induce large matter density fluctuations at horizon re-entry, which can collapse to form PBHs.

At leading order in slow-roll, the amplitude of the scalar power spectrum in single field inflation with canonical kinetic terms reads
\beq
\Delta_s^2\,=\,\frac{1}{4 \pi^2}\,
\frac{H^4}{|\dot \phi|^2}\bigg\rvert_{k=aH} \,, \label{leare1}
\eeq
where $H$ is the Hubble parameter, and $\dot \phi$ denotes the time derivative of the inflaton. Deviations from a slow-roll regime  can change this expression, but at first sight it seems  hard to increase its value by several orders of magnitude within a well-defined  range of scales, without spoiling inflation. On the other hand, various recent works have succeeded in doing so \cite{Garcia-Bellido:2017mdw}, by using inflationary potentials with inflection points. In the inflection point regions of the potential the inflaton dynamics experiences a rapid speed decrease, and enters into a so-called ultra slow-roll regime  during which  the amplitude of $\Delta_s^2$ can indeed be enhanced by several orders of magnitude.

A heuristic, physically transparent explanation for this phenomenon can be found in works by Leach and Liddle \cite{Leach:2000yw} and Leach et al. \cite{Leach:2001zf}. We review their argument here, using it as guideline for the discussion we develop in the remaining sections.

 The mode equation for curvature fluctuations of wavenumber $k$ in single field inflation reads 
 \beq\label{mez1}
\mathcal{R}''_k + 2\fr{z'}{z}\mathcal{R}'_k
 + k^2
\mathcal{R}_k = 0, 
\eeq
where the ``pump field'' for the curvature perturbation defined as $z \equiv a \dot{\phi}/H$ satisfies:
\beq
\fr{z'}{z}= aH \left(1+\epsilon-\delta \right). \label{defzpz1}
\eeq
In this expression, the standard slow-roll parameters $\epsilon$ and $\delta$ are defined as
\begin{eqnarray}
\epsilon &\equiv& -\frac{\dot H}{H^2}\,,
\\
\delta &\equiv& -\frac{\ddot H}{2H\dot H} \,=\,-\frac{\dot \epsilon}{2\,\epsilon\,H}+ 
\epsilon\,.
\end{eqnarray}
At a given moment of time during inflation, $\epsilon<1$, while $\delta$ can be in principle of any size. 
Notice that the mode equation in \eqref{mez1} is of the form of a damped harmonic oscillator. In the standard slow-roll limit $\epsilon,\delta \ll 1$, we focus on modes $\mathcal{R}_k$ which already left the horizon (\ie $k < |z'/z|$).

The  solution to equation \eqref{mez1} can then be
expressed as
\beq \label{solgeq1}
\mathcal{R}_k(\tau)\,=\,{\cal C}_1+{\cal C}_2\,\int \frac{\d \tau}{z^2}
\eeq
where ${\cal C}_1$ and ${\cal C}_2$ are two integration constants,  multiplying respectively the constant and 
decaying solutions to the equation \eqref{mez1}. Since  equation \eqref{defzpz1} implies
\beq \label{solzta1}
z(a)\,=\,z_0\,\exp{\left[\int\,\left(1+\epsilon-\delta\right)\,\d\ln{a} \right]}
\eeq
with $z_0$ a constant, we learn that  $z \sim a$  in the limit where the slow-roll parameters $\epsilon$ and $\delta$
can be neglected. Since in such limit $a\sim-1/(H \tau)$, the decaying mode proportional to  ${\cal C}_2$ rapidly 
decays as $a^{-3}$ outside the horizon,  and the curvature perturbation is then conserved at super-horizon scales,  being controlled by the constant mode ${\cal C}_1$. After matching with the  Bunch-Davies vacuum at sub-horizon scales,  one obtains the power spectrum amplitude of eq.~\eqref{leare1} at leading order in a slow-roll approximation.

Departure from the slow-roll regime suggests a way to enhance the amplitude of curvature perturbations for modes at certain scales, right after
 they cross the horizon.
  Suppose that the friction term proportional to $z'/z$ in \eqref{mez1}
   transiently 
   {\it changes sign} for
  some short interval
   during the inflationary homogeneous evolution: 
\beq \label{condtste1}
1+\epsilon - \delta < 0\,.
\eeq 
 In this case,  the friction becomes a driving term in the equation \eqref{mez1}: the exponent in eq \eqref{solzta1} becomes negative, hence $z$ decreases with time instead
of increasing,  implying that the mode proportional to ${\cal C}_2$  appearing in  eq \eqref{solgeq1} is  growing in this regime. Its contribution to the curvature perturbation can
become substantial, increasing the size of ${\cal R}$.  The once-decaying, now-growing mode ${\cal C}_2$
  can thus be exploited to enhance the power spectrum of curvature fluctuations during a short range of scales
 \cite{Leach:2001zf}.
Since $\epsilon$ is positive and at most order one, the condition \eqref{condtste1} requires $\delta$ to be at least order one, implying that this can occur through a transition to fast-roll (where $\delta =1$), during which the slow-roll approximation breaks down.  In particular, the ultra slow-roll regime mentioned above (and that we shall discuss more at length later on) corresponds to a phase during which $\delta \ge 3$.

 This is the basic physical mechanism 
 we  intend to exploit to enhance the power spectrum at small scales, and thus produce PBHs. We will apply and generalise  it 
 to two string motivated scenarios for single field inflation, the first being based on  axion inflation, the second D-brane
 inflation with non-canonical kinetic terms. 
   
\section{PBHs  in axion inflation with  subleading non-perturbative effects}\label{Sec2}

\bigskip

In this section we analyse in full detail a string motivated model for axion  inflation, which includes next-to-leading,
non-perturbative contributions to a monomial potential.  We  proceed as follows
\begin{itemize}
\item We start by presenting the theoretical motivations underlying this system.  We show how subleading,
non-perturbative corrections to the axion potential can qualitatively alter the homogeneous dynamics
of the inflaton field, and work out the corresponding time evolution of the slow-roll parameters (Sections \ref{sec-bum}-\ref{S:ultraslow}).
\item We continue in Sections \ref{cmb-sec}-\ref{EPSPBH} by numerically
  studying the dynamics of curvature fluctuations in two concrete
models based on this set-up, showing that they can be in good agreement with CMB measurements (typically predicting a large value for the running parameter $\alpha_s$), and at the same time produce
an enhancement of the curvature power spectrum at small scales, exploiting the argument of Section \ref{SecHeu}.
\item In Section \ref{PBH} we then show
that our models produce a monochromatic population of light PBHs that can provide a considerable fraction of Dark Matter density. We discuss
 observational constraints on our PBH features, 
 and further
constraints that the production mechanism imposes on the cosmological evolution after inflation ends. 
\end{itemize}



\subsection{Bumpy inflation}\label{sec-bum}

We  consider a scenario that is based upon an earlier observation on how subleading, non-perturbative effects can alter axion inflation \cite{Parameswaran:2016qqq}.  In string axion inflation, the perturbative axion shift symmetry is broken spontaneously by background vevs (\eg fluxes) or non-perturbative effects (\eg string instantons), leading to large field inflation models with monomial or cosine (``natural inflation'') potentials.  In \cite{Parameswaran:2016qqq}, we noted that subleading non-perturbative corrections -- if sufficiently large -- can superimpose oscillations onto the underlying potential. The size of these effects will depend on the vev's of fluxes and other moduli, which are already stabilised. Therefore, they may be  tiny,  large enough to introduce new local minima and maxima that may halt inflation, or anything in between. 
We focus on an intermediate situation, where step-like features are induced in the potential, with steep cliffs and gentle plateaus, which transiently induce large deviations from the slow-roll attractor regime.

\smallskip

For concreteness, we  consider a string-inspired model with axion, $\phi$, with a canonical kinetic term and minimal coupling to gravity:
\beq\label{L}
\fr{\mathcal{L}}{\sqrt{-g}} = \fr{\Mp^2}{2}R - \fr{1}{2}\partial_{\mu}\phi\partial^{\mu}\phi-V(\phi),
\eeq 
where the axion potential takes the following form
\beq\label{pot}
V(\phi) = V_0 + \fr{1}{2}m^2\phi^2 + \Lambda_1^4~ \fr{\phi}{f}~\cos\left(\fr{\phi}{f}\right)+ \Lambda_2^4 \sin\left(\fr{\phi}{f}\right).
\eeq
This class of potentials is known to arise from string theory constructions  \cite{Kobayashi:2015aaa,BLZ,Kallosh:2014vja,Flauger:2014ana}\footnote{For example, the potential of the form above arises for an axion $Im(Z)$ - after having fixed the saxion $Re(Z)$ -  from a K\"ahler potential $K = - \ln(Z + \bar Z)$, superpotential $W = W_0 + M Z + i \Lambda e^{-bZ}$ and an uplift term, motivated e.g. by fluxes and non-perturbative effects. The coefficients in the potential will then depend on the fluxes $W_0, M, \Lambda$, as well as the vev of the saxion.}.

The background dynamics of the inflaton depends on the size of the non-perturbative corrections compared to the mass term in the potential \eqref{pot}, in particular on the ratios  $\beta_i \equiv \Lambda_i^4/m^2 f^2$. In the limit $\beta_i \to 0$ ($i=1,2$), non-perturbative corrections become negligible and we recover the usual smooth quadratic potential.  For $\beta_i > 1$, one introduces a large number of new stationary points (where $V'= 0$) into the smooth $\phi^2$ potential in a given range of field values. In this case, the classically rolling scalar field might eventually get stuck into  some local minimum depending on the initial conditions \cite{Banks:2003sx}.  In this work, we  focus on the parameter space where $\beta_i < 1$  for both $i=1,2$, but without assuming $\beta_i \ll 1$. 

\begin{figure}[t!]
\begin{center}
\includegraphics[scale=0.65]{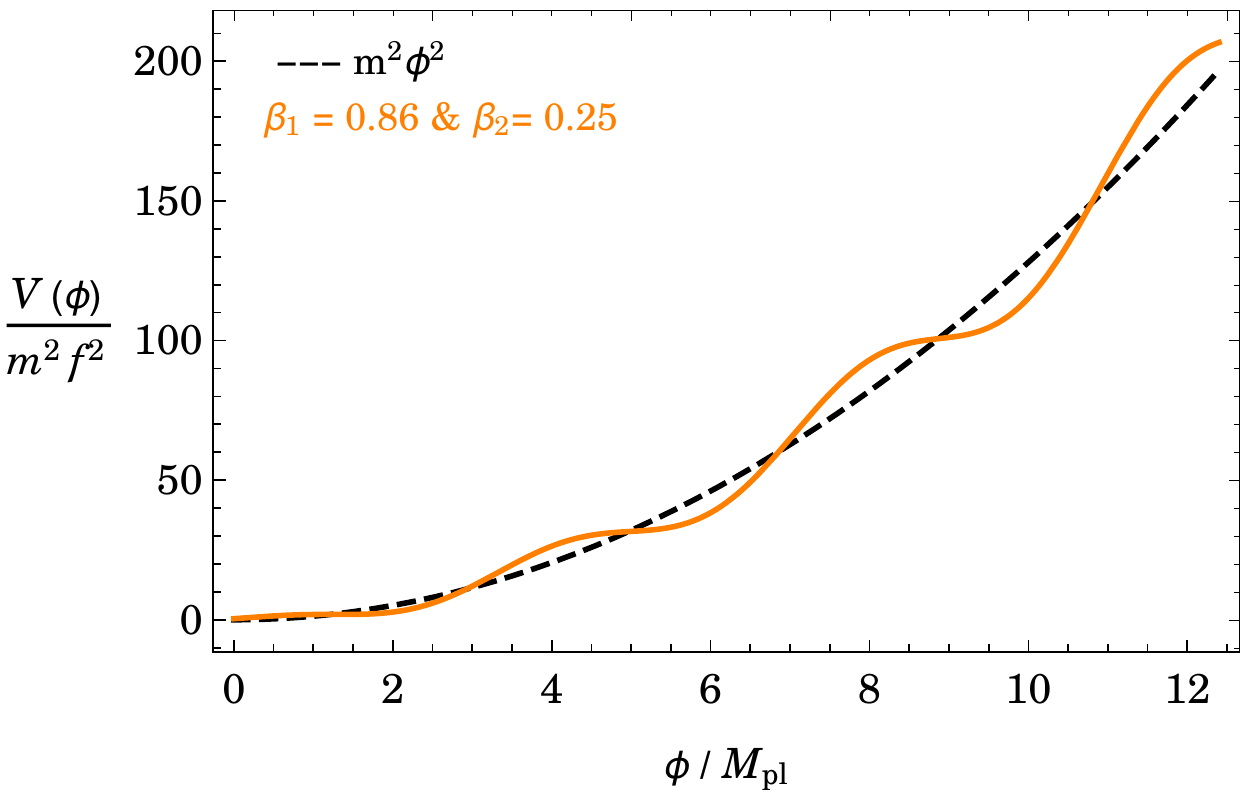}\includegraphics[scale=0.64]{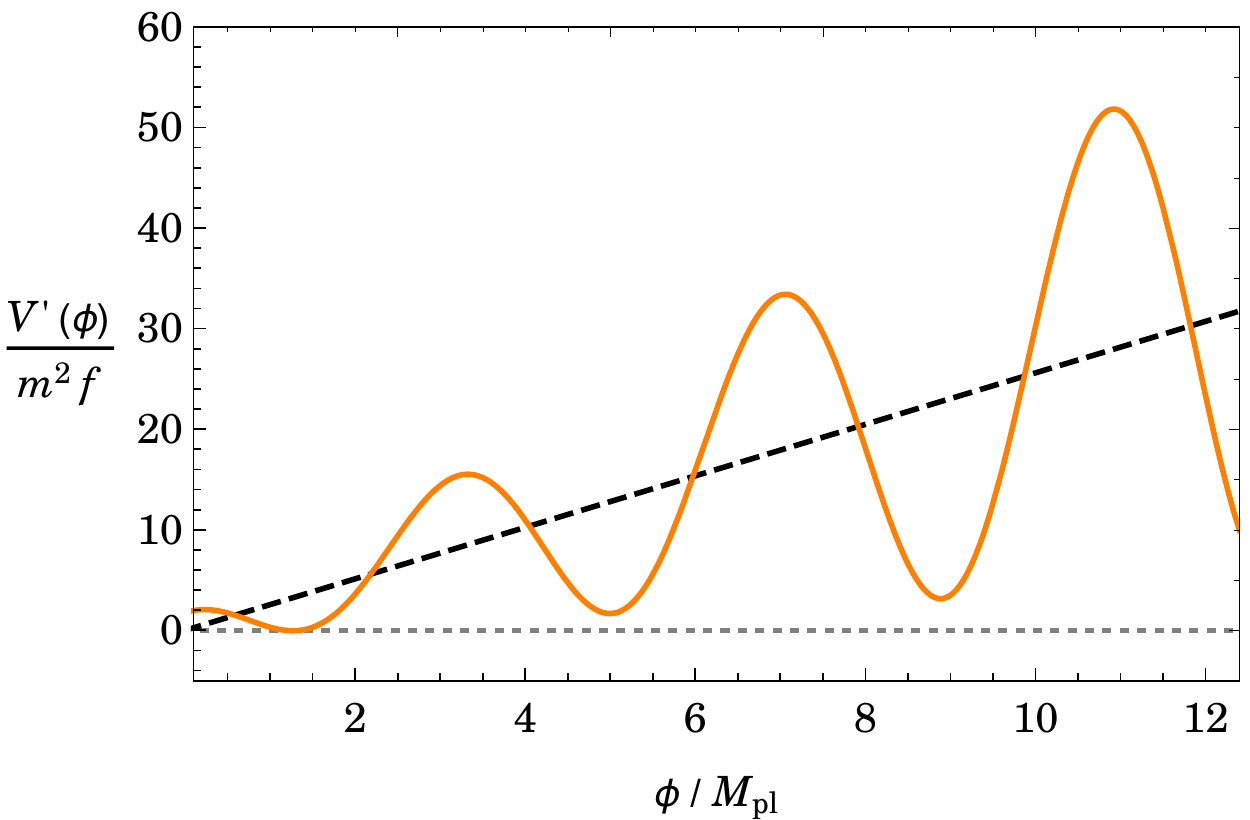}
\end{center}
\caption{Potential $V(\phi)$ (left, orange) in \eqref{pot} and its derivative $V'(\phi)$ (right, orange) for parameters $\beta_1 
\equiv \Lambda_1^4/m^2f^2 = 0.86$, $\beta_2 
\equiv \Lambda_2^4/m^2f^2 = 0.25$ and $\Mp/f = 1.6$, in comparison to the case of smooth quadratic potential $V_{sm}(\phi)\propto \phi^2$ (black, dashed). The gray dotted line on the right plot is shown to guide the eye towards $V'(\phi) = 0$.\label{fig:V}}
\end{figure}

To illustrate the general shape of the potential we are interested in, in Figure \ref{fig:V} we plot  $V(\phi)$ in \eqref{pot} and its slope for the parameters 
\beq
\beta_1 \simeq 0.86\,,\hskip1cm \beta_2 \simeq 0.25\,,\hskip1cm \Mp/f = 1.6 \,,\hskip1cm 
{\text{(Case 1)}}
\label{cca1}
\eeq
while $V_0$ is chosen to ensure that the potential is vanishing at the minimum.
   The non-perturbative corrections, being subleading but considerable, introduce plateau-like regions connected by steep cliffs. Notice that the slope of the potential, $V'$, is positive for a large range of $\phi$ values but gradually decreases until it eventually vanishes at a shallow local minimum when $\phi \sim 1.35 \Mp$.  One can expect the dynamics to be such that an initially displaced $\phi$  rolls down in its wiggly potential, passing through the local minimum,  and eventually settling on its global minimum at $\phi=0$ \cite{Parameswaran:2016qqq}. In the upcoming sections, we will elaborate on the interesting dynamics that arises due to the presence of a shallow local minimum, shortly before the global minimum.

\subsection{Background evolution -- slow roll, fast roll}\label{BBack}
\begin{figure}[t!]
\begin{center}
\includegraphics[scale=0.64]{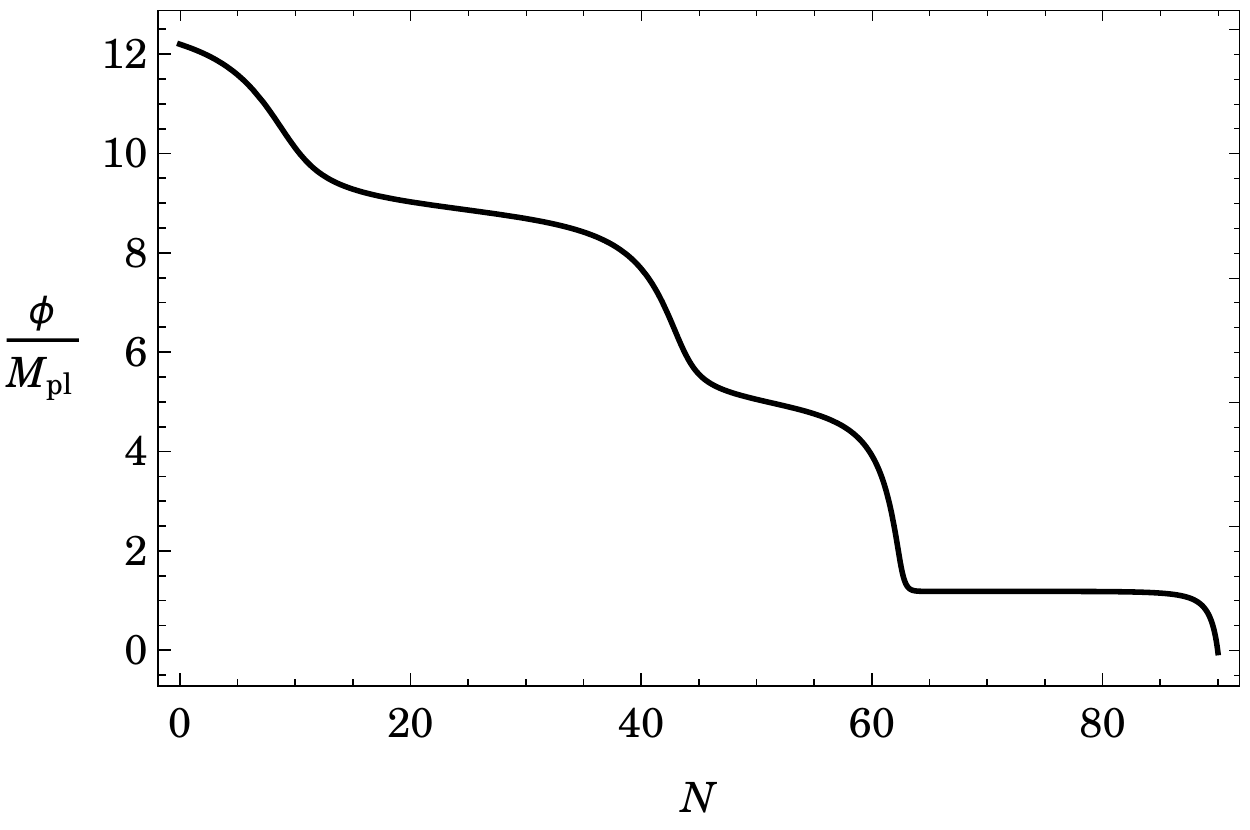}~~\includegraphics[scale=0.617]{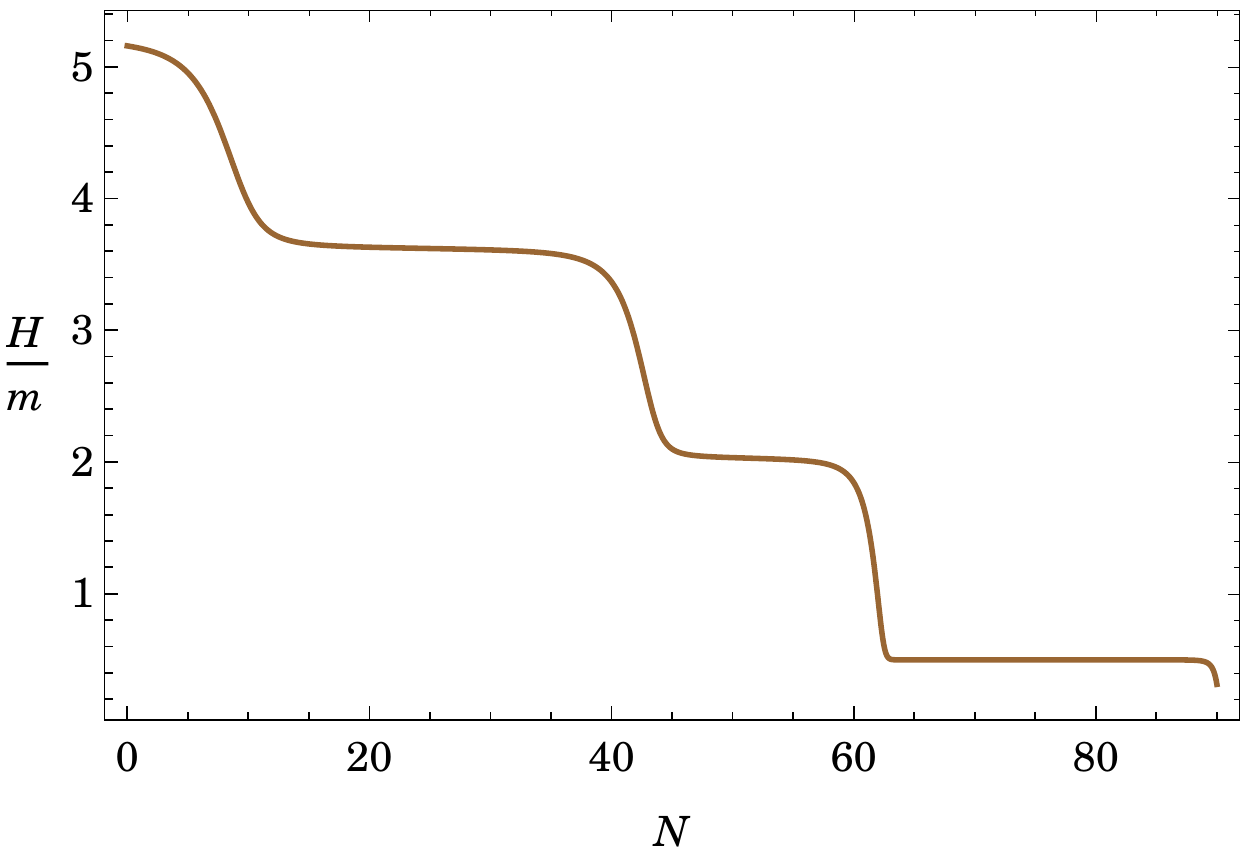}
\end{center}
\caption{Background solution  $\phi(N)$ (left) and $H(N)/m$ (right) in the bumpy potential \eqref{pot} with the initial condition $\phi(0) = 12.2~ \Mp$ where $\beta_1$, $\beta_2$ and $\Mp /f$ are taken to be the same as in Figure \ref{fig:V}. \label{fig:Back}}
\end{figure}

We now study the inflationary dynamics that arises from the Lagrangian \eqref{L} and 
\eqref{pot} on a flat FRW background.  Using the number of e-folds, $N(t) = \ln a(t)$, as the time variable, the system is governed by the following set of equations:
\bea\label{FRW}
\nn H^2 &=& \fr{V(\phi)}{\Mp^2(3-\epsilon)},\\
\fr{\d^2 \phi}{\d N^2} &+& \left(3 - \epsilon\right)\fr{\d\phi}{\d N} + \fr{1}{H^2}V'(\phi) = 0,
\eea
where $\epsilon$ is the standard Hubble slow-roll parameter,
\beq\label{eps}
\epsilon = -\frac{\dot H}{H^2} = \fr{1}{2\Mp^2} \left(\fr{\d \phi}{\d N}\right)^2.
\eeq 
We numerically solve the set of equations \eqref{FRW} and \eqref{eps} assuming initially we are in the slow-roll attractor regime, defined by the condition
\beq
\fr{\d \phi}{\d N} = -\fr{V'(\phi)}{V(\phi)}.
\eeq
In this way, we set all the initial conditions required to solve the system only using a given initial $\phi$ value. As an example, we set $\phi(N_{\rm in}=0)= 12.2 \Mp$,  and plot the solutions to \eqref{FRW} as a function of e-folds during inflation in Figure \ref{fig:Back}. 

We find  that the inflaton slowly rolls down the smooth plateau-like regions, sustaining an almost constant Hubble friction.  However, whenever it meets a cliff, the inflaton speeds up quickly, until it reaches the next plateau where Hubble friction rapidly  slows it right back down again \cite{Parameswaran:2016qqq}.   Notice also that the strength of the acceleration down the cliffs increases as the field rolls down to small field values.  The system is in a slow-roll attractor regime within the plateaus, but departs from slow-roll during the acceleration and fast roll through the steeper cliffs and the during the deceleration when rolling into the flat plateaus from the steep cliffs. 

\begin{figure}[t!]
\begin{center}
\includegraphics[scale=0.615]{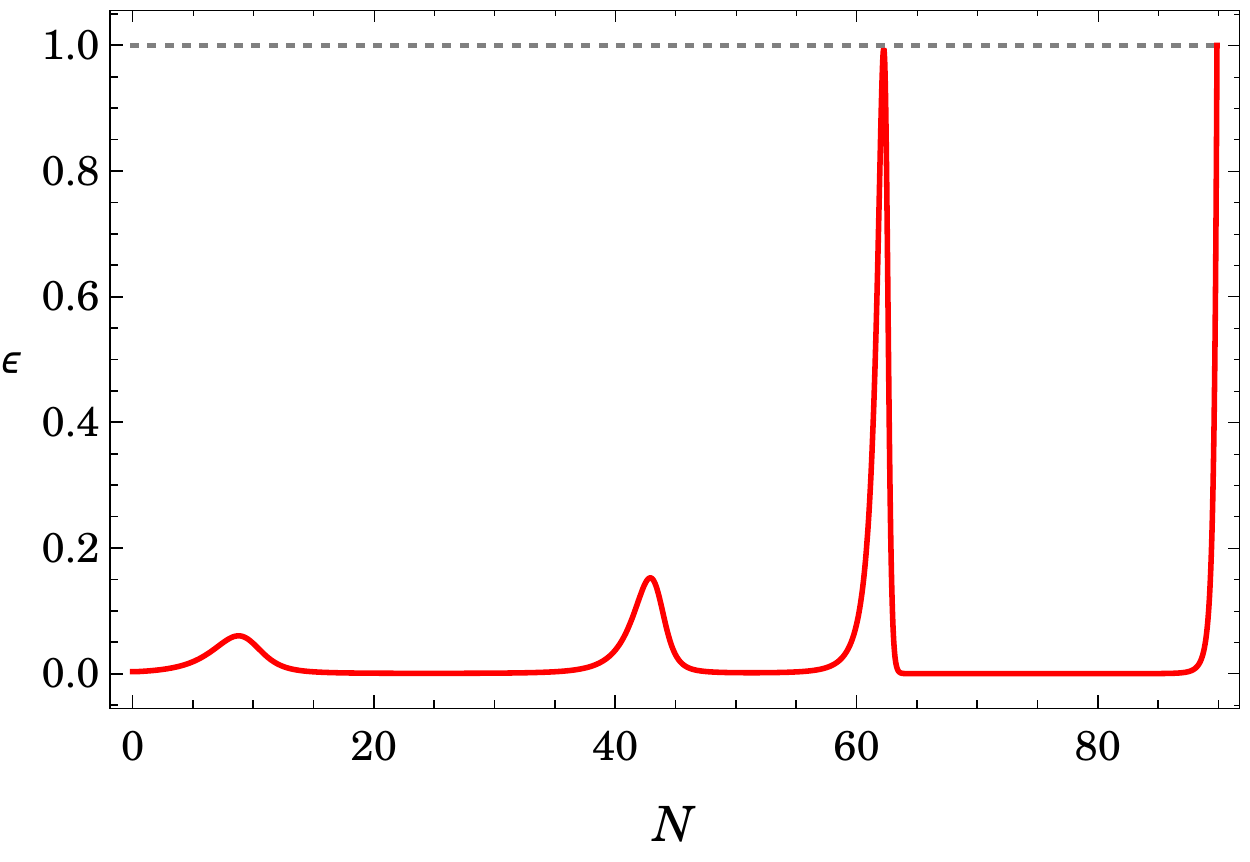}~~\includegraphics[scale=0.6]{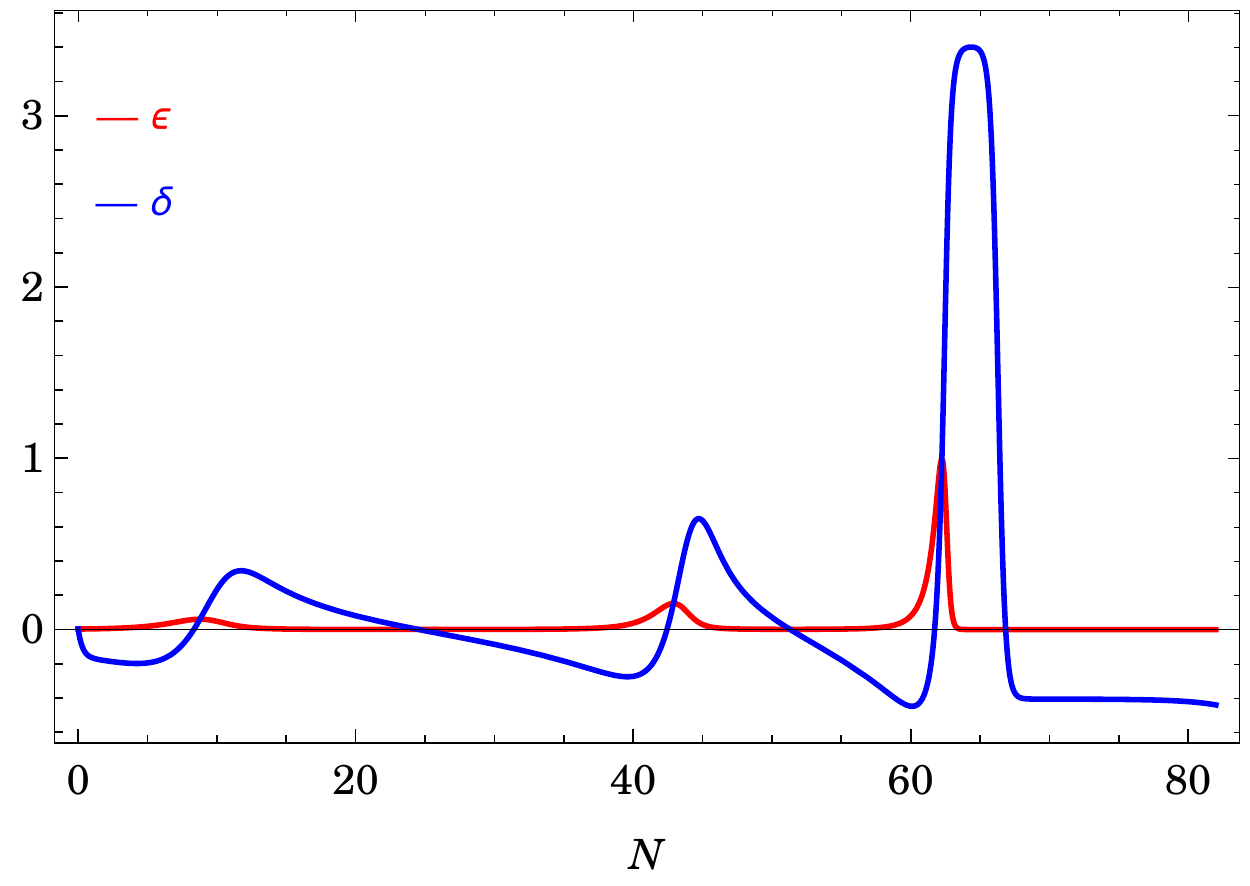}
\end{center}
\caption{Evolution of the Hubble slow-roll parameters $\epsilon$ (left) and $\delta$ together with $\epsilon$ (right) as a function of e-folds $N$. \label{fig:eps}}
\end{figure}

This behaviour can be seen from Figure \ref{fig:eps} where we plot the evolution of the slow-roll parameter $\epsilon$ together with the parameter:
\beq
\delta \equiv -\frac{\ddot H}{2H\dot H} 
= - \frac{\ddot{\phi}}{\dot{\phi} H}
\,,
\eeq
with respect to e-folds $N$. We see peaks in the slow-roll parameter $\epsilon$ and oscillations in $\delta$ as $\phi$ accelerates ($\delta < 0$)  down the steep cliffs and decelerates ($\delta >0$) into the plateaus\footnote{Note that in our conventions,
 the field rolls down from large to small values: this implies that the velocity is
 $\dot{\phi} < 0$, and  acceleration corresponds  to  $\ddot{\phi} < 0$. 
 Hence acceleration (deceleration) of the inflaton corresponds to the case where $\delta < 0$ ($\delta > 0$).}.  

Note that although $\epsilon$ increases quite considerably at around $N\approx 62$, 
we still have $\epsilon < 1$, hence
the accelerated spacetime expansion does not terminate here.  Inflation lasts until $N_{\rm tot} \simeq 90	$ e-folds\footnote{Note that only the last $\sim 65$ e-folds or so of inflation have observational consequences.} where $\epsilon= 1$, and $\phi$ begins to settle at its global minimum.  

Parameter choices for the axion potential \eqref{pot} are also possible such that $\phi$ rolls down the last cliff so fast that $\epsilon$ becomes larger than unity for a short period of time -- temporarily halting inflation -- before it rapidly decelerates at the plateau, and resumes inflation.  In Figures 
\ref{fig:BackIS},\ref{fig:epsIS} and \ref{fig:eps1IS}, we present such a model, numerically solving \eqref{FRW} with parameters
\beq
\beta_1 \simeq 0.9\,,\hskip1cm \beta_2 \simeq 0.16\,,\hskip1cm \Mp/f = 1.7 \,,\hskip1cm 
{\text{(Case 2)}} \label{cca2}
\eeq 
and the initial condition $\phi(0)= 11.55~ \Mp$. 

\subsection{Ultra slow-roll phase} \label{S:ultraslow}

\begin{figure}[t!]
\begin{center}
\includegraphics[scale=0.645]{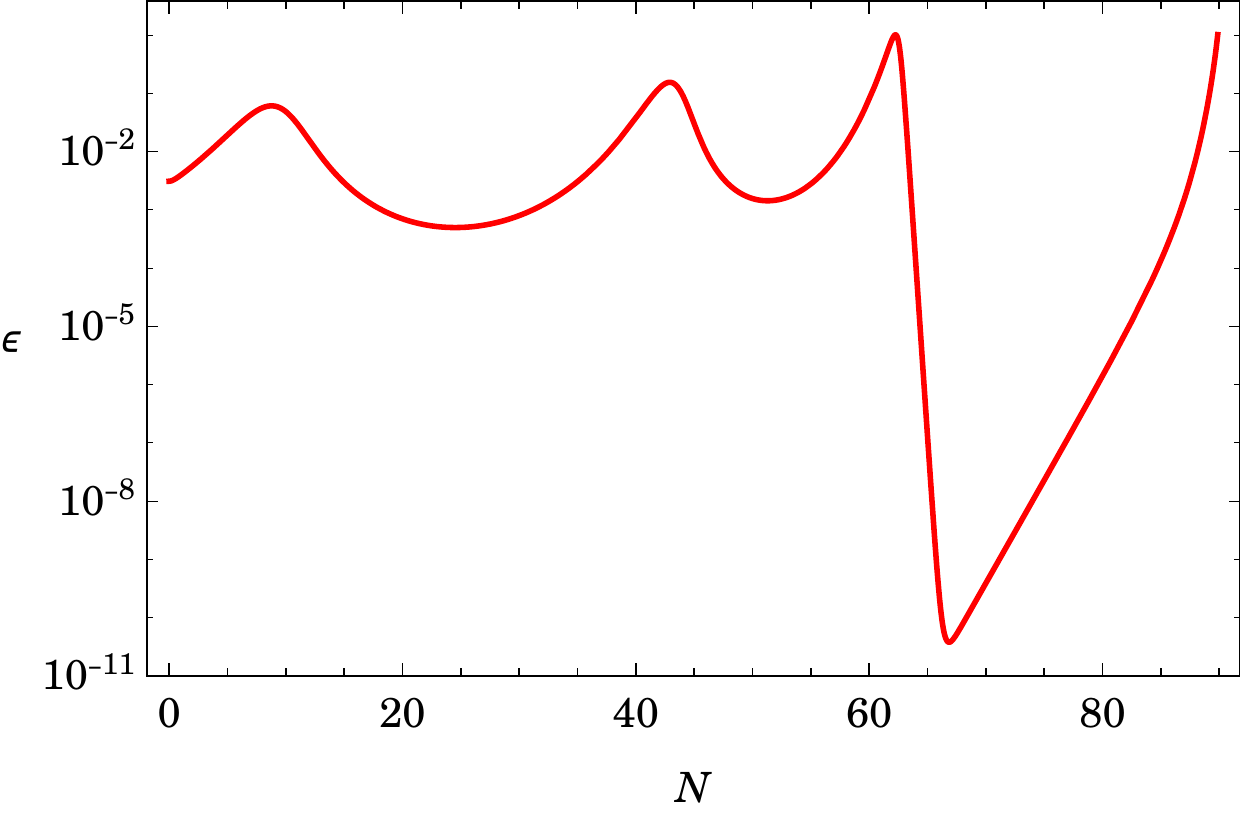}
\includegraphics[scale=0.605]{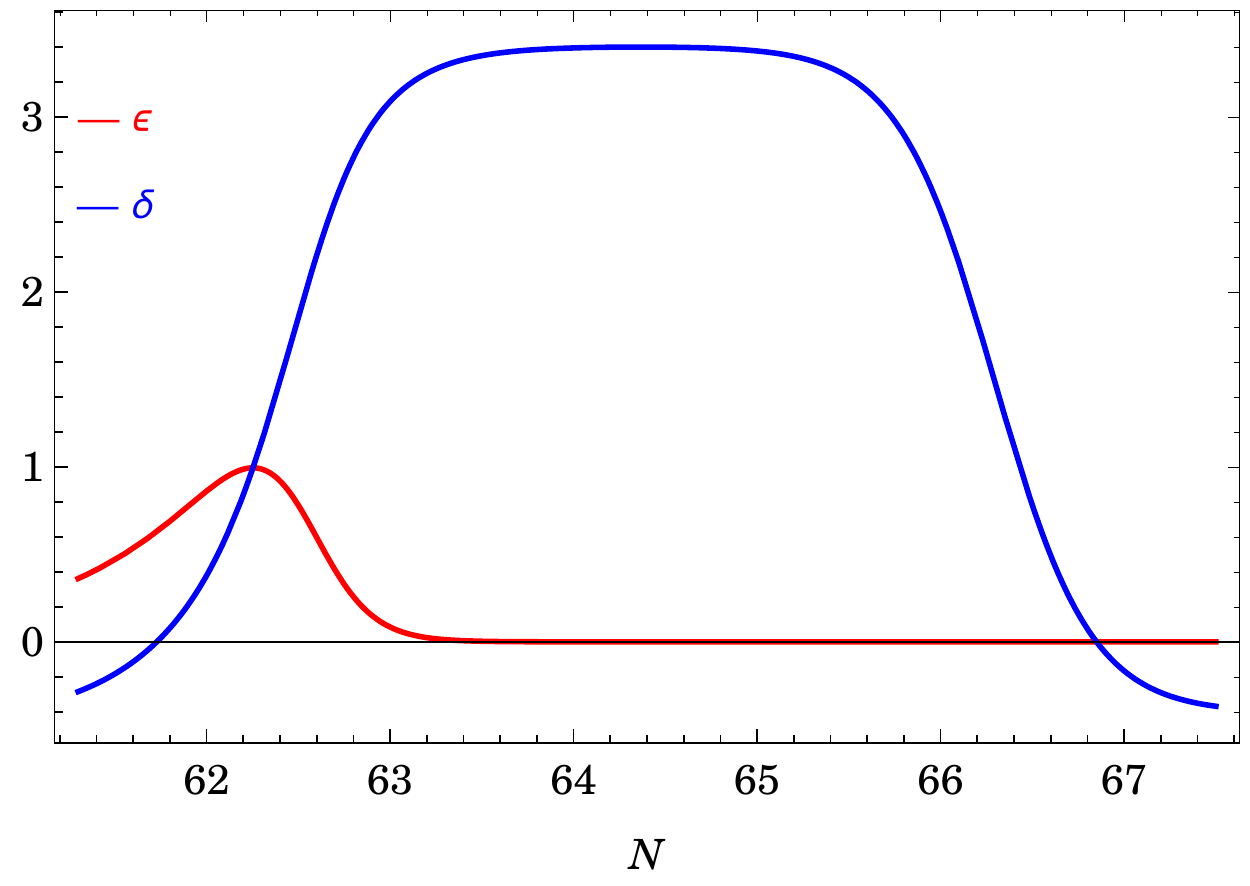}
\end{center}
\caption{Evolution of $\epsilon$ in a logarithmic scale as a function of e-folds (left). Zoomed in plot for $\delta$ together with $\epsilon$ around the time where $\delta > 1$ corresponding to the evolution of the inflaton $\phi$ after the last cliff in the potential \eqref{pot} (See \eg, Figure \ref{fig:zV}).\label{fig:eps1}}
\end{figure}

\begin{figure}[t!]
\begin{center}
\includegraphics[scale=0.65]{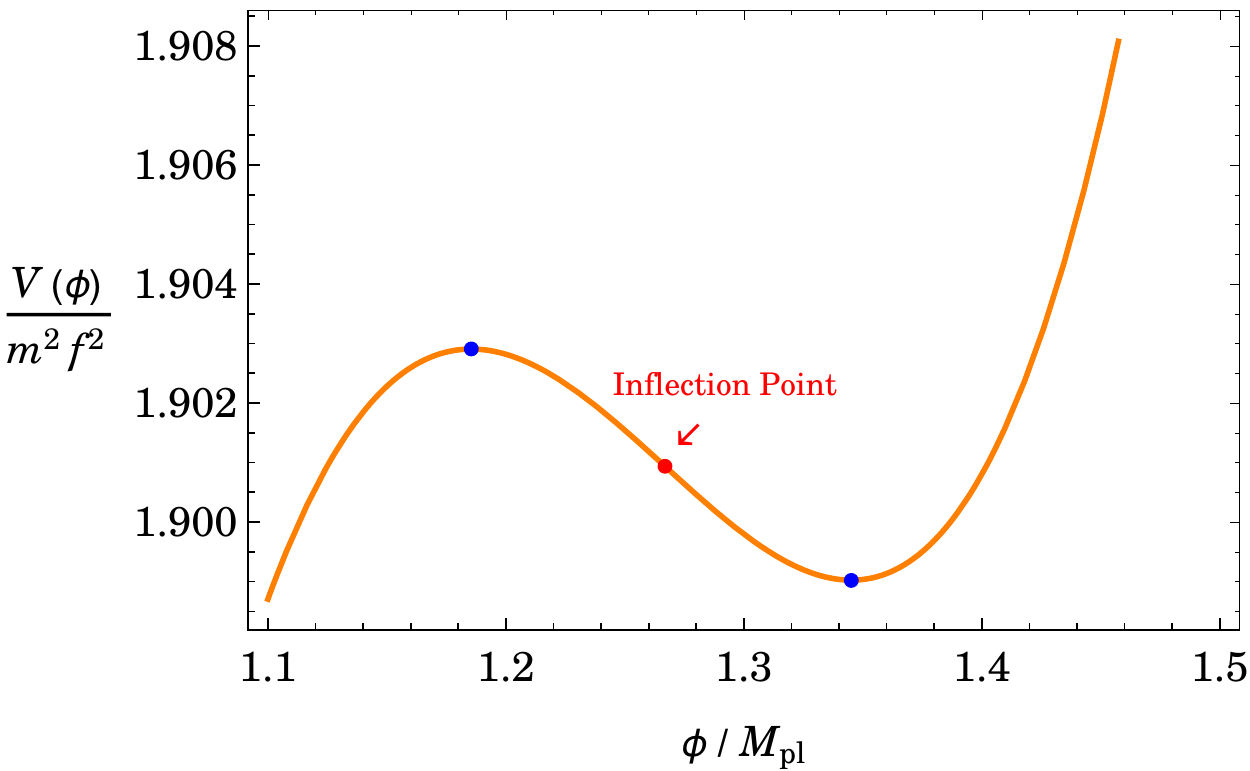}
\includegraphics[scale=0.61]{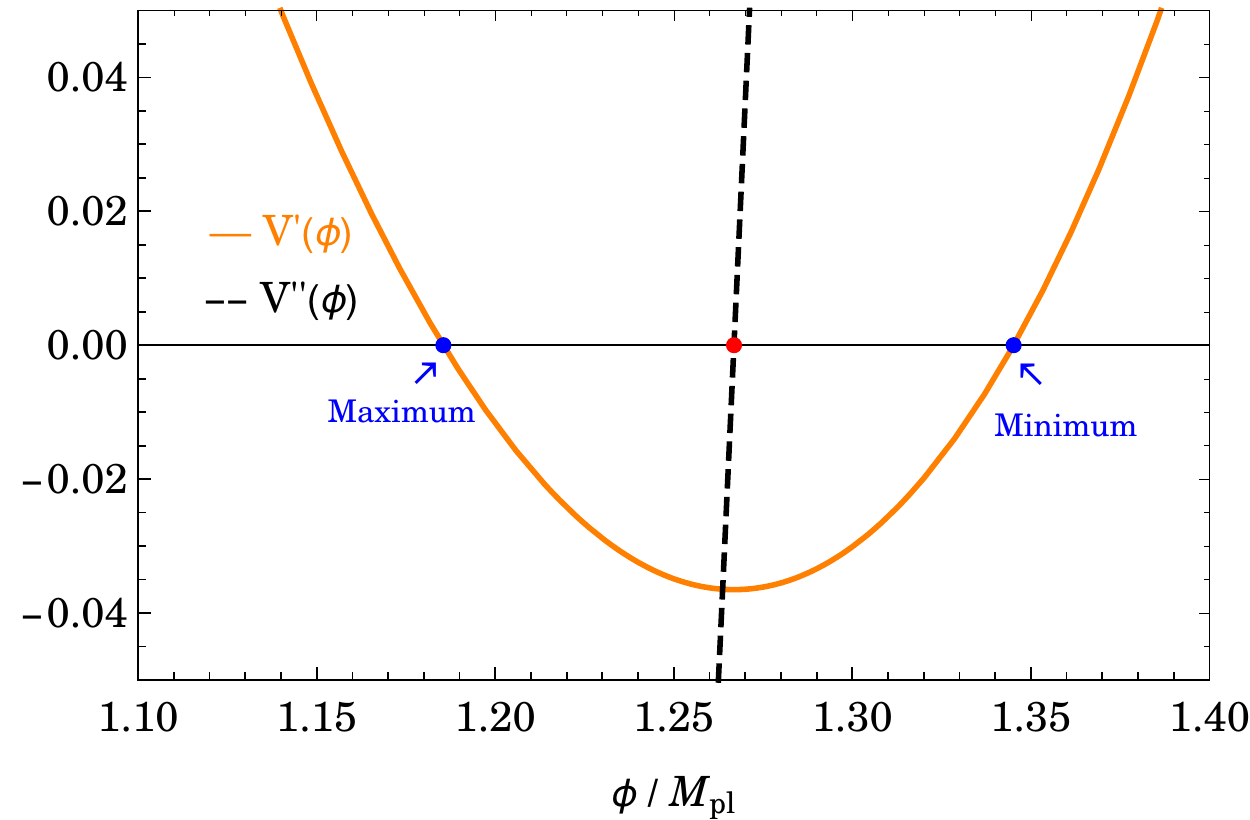}
\end{center}
\caption{The shape of the potential $V(\phi)$ in \eqref{pot} after the last cliff (left) and the behaviour of $V'(\phi)$ and $V''(\phi)$ in the same field range (right) with the same parameter choices as in Figure \ref{fig:V}.    Blue points in both graphs represents the points where $V'= 0$ whereas the red dot is an inflection point where $V''=0$. \label{fig:zV}} 
\end{figure}

\begin{figure}[t!]
\begin{center}
\includegraphics[scale=0.63]{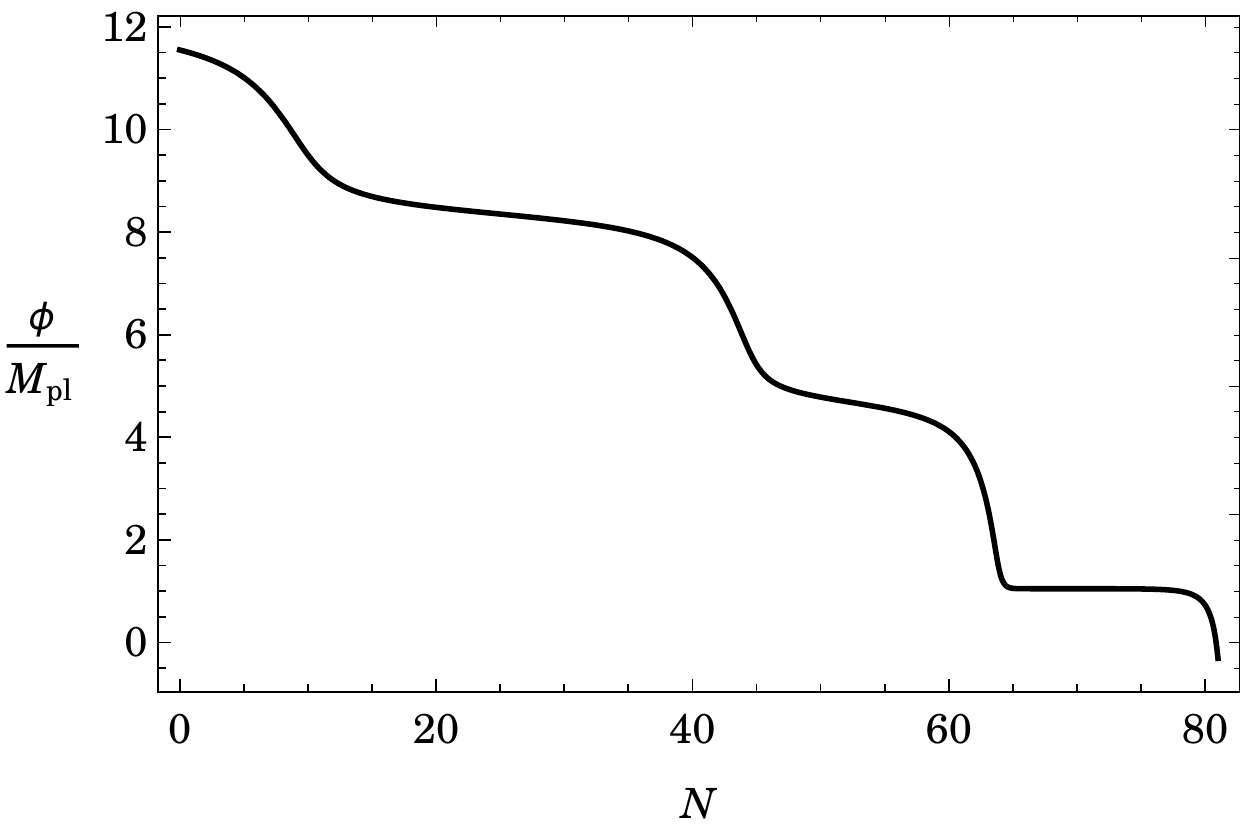}~~\includegraphics[scale=0.605]{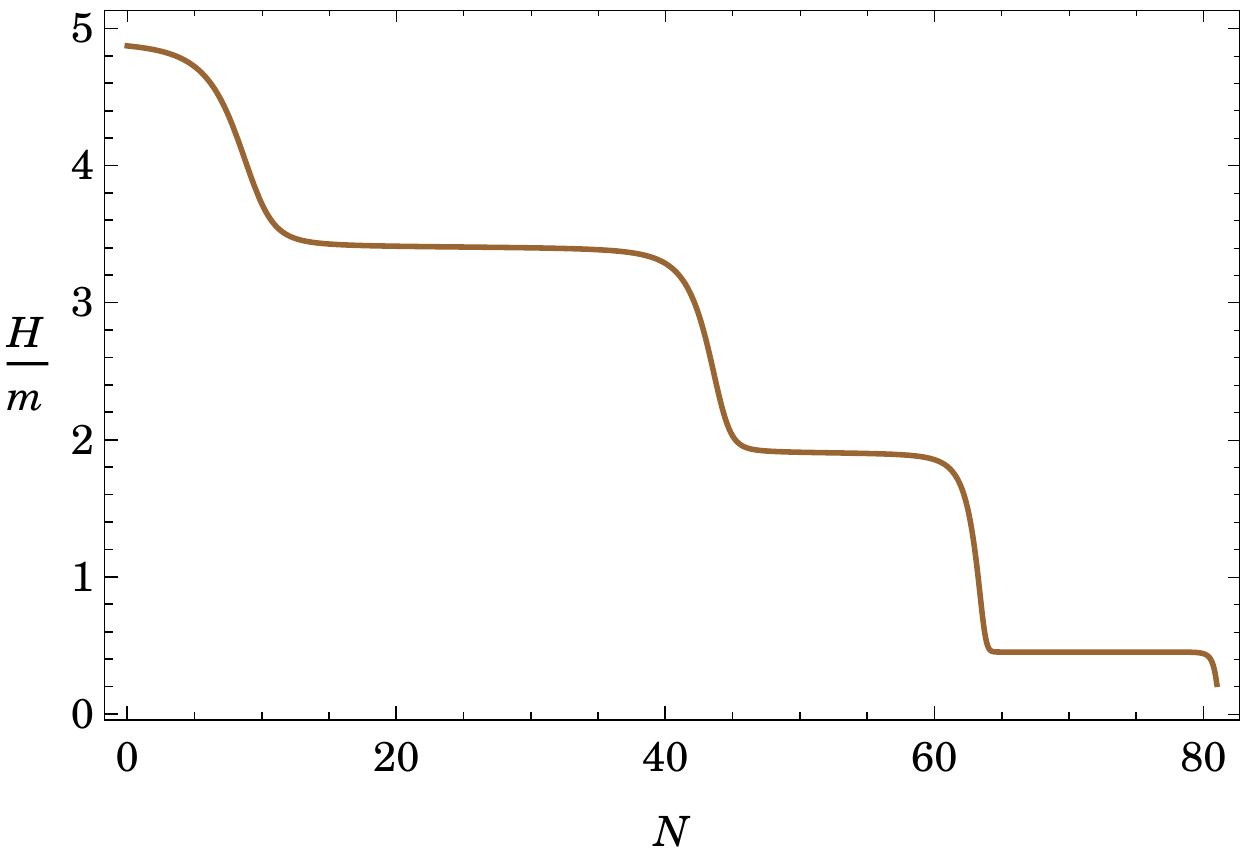}
\end{center}
\caption{Background solution to $\phi(N)$ (left) and $H(N)/m$ (right) in the bumpy potential \eqref{pot} with the initial condition $\phi(0) = 11.55~ \Mp$ where $\beta_1 \simeq 0.9$, $\beta_2\simeq 0.16$ and $\Mp /f=1.7$. \label{fig:BackIS}}
\end{figure}

\begin{figure}[t!]
\begin{center}
\includegraphics[scale=0.6]{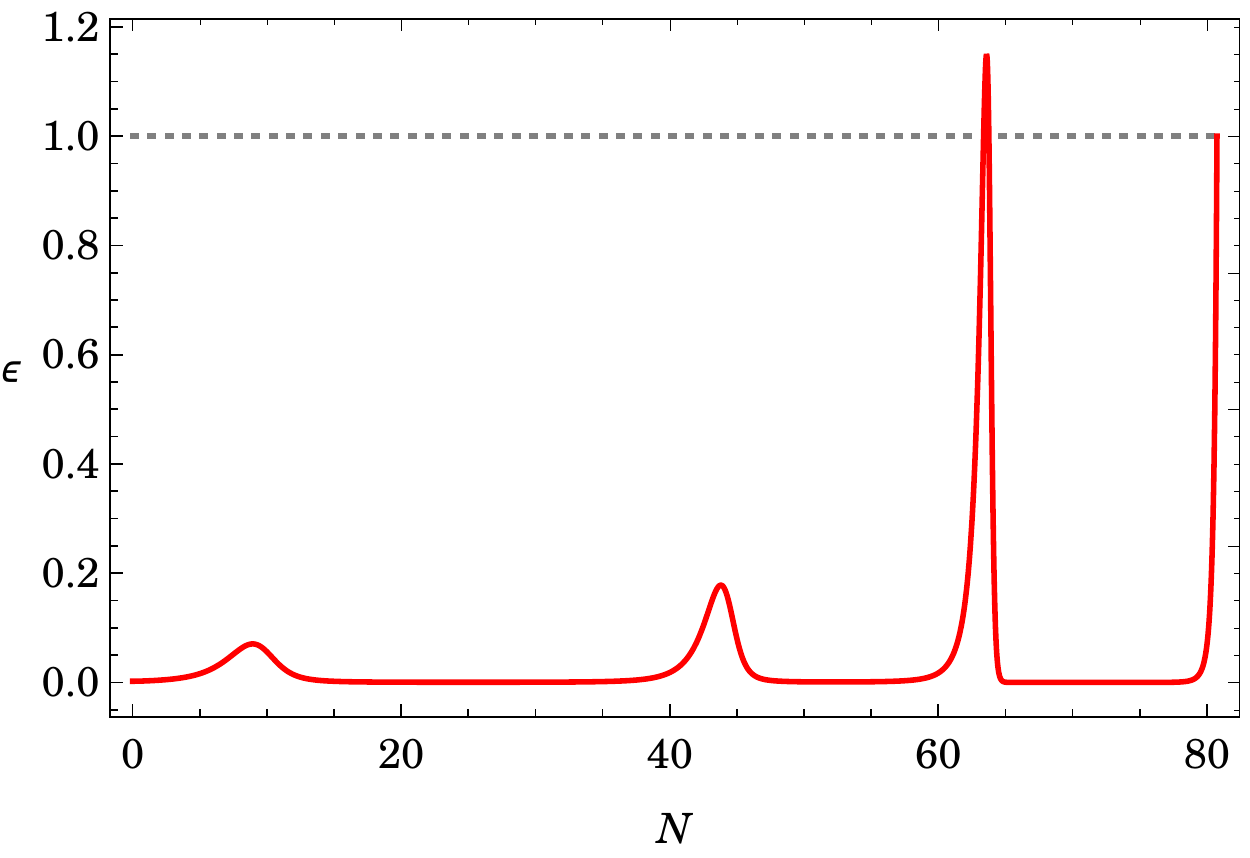}~~\includegraphics[scale=0.587]{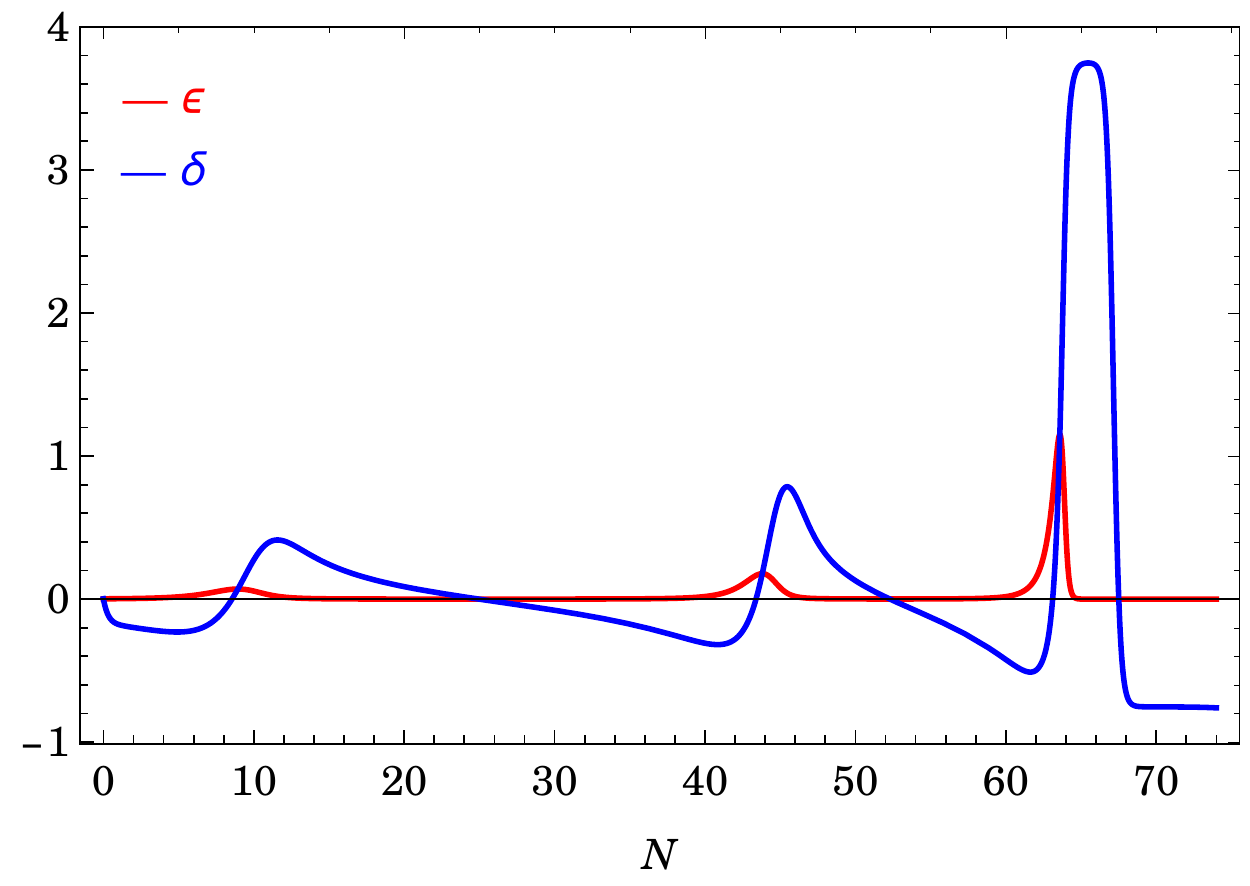}
\end{center}
\caption{Evolution of the Hubble slow-roll parameters $\epsilon$ (left) and $\delta$ together with $\epsilon$ (right) as a function of e-folds $N$ for the parameter choices as in Figure \ref{fig:BackIS}. \label{fig:epsIS}}
\end{figure}

\begin{figure}[t!]
\begin{center}
\includegraphics[scale=0.61]{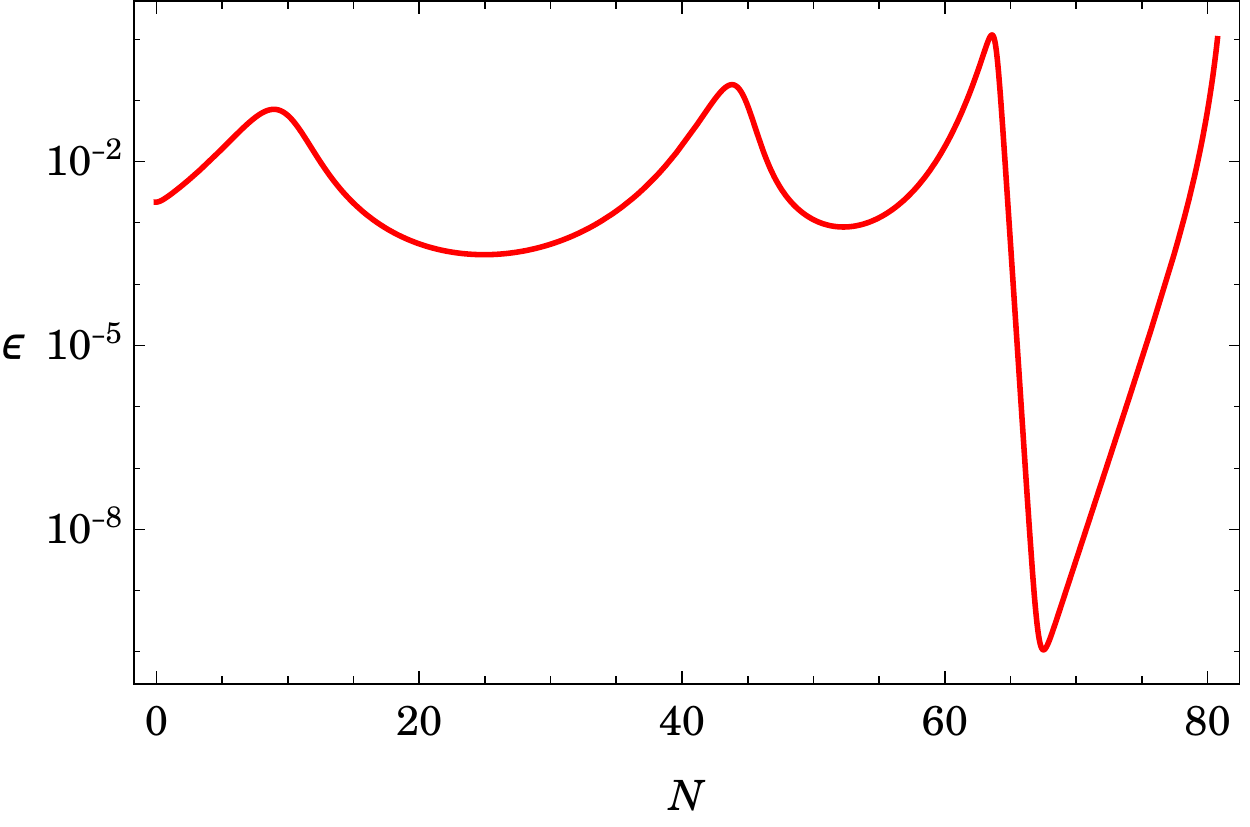}~~
\includegraphics[scale=0.58]{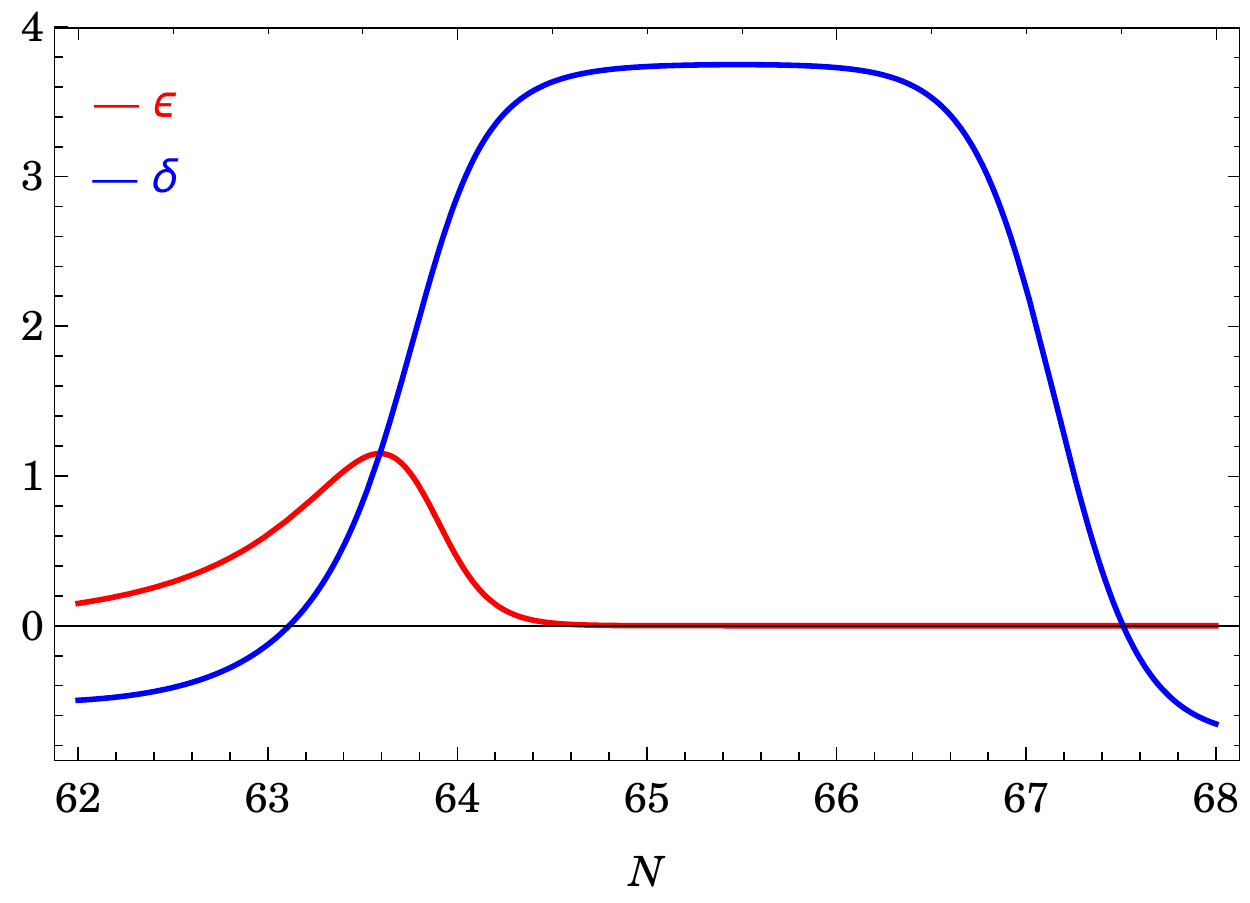}
\end{center}
\caption{Evolution of $\epsilon$ in a logarithmic scale as a function of e-folds (left). Zoomed in plot for $\delta$ together with $\epsilon$.\label{fig:eps1IS}}
\end{figure}

The most important feature of the bumpy potential \eqref{pot} is the behaviour of $\phi$ after the last cliff and through the local minimum towards the global one. Let us discuss the Case 1 parameters in eq \eqref{cca1}, as Case 2 is very similar.  As shown in Figure \ref{fig:eps1}, at around $N \gtrsim 62$, the inflaton starts to decelerate enormously,  while the slow-roll condition, $\delta < 1$ is violated for about $\Delta N \approx 4$ e-folds during which $\delta$ reaches  a maximum value of\footnote{By definition, $\delta \equiv 3 + V'(\phi)/(\dot{\phi} H)$ using the KG equation in cosmological time. This explains why $\delta > 3$ during ultra slow-roll as $V'<0$ around the inflection point (See \eg, Figure \ref{fig:zV}).}  $\delta \gtrsim 3$. During the time where $\delta > 1$, the friction term in the Klein-Gordon (KG) equation is not balanced by the slope of the potential as in the standard slow-roll attractor case but with the acceleration term in \eqref{FRW}, and shortly after the system enters an `ultra slow-roll' regime where $\epsilon \ll 1$ and $\delta \gtrsim 3$ \cite{Kinney:2005vj,Martin:2012pe,Germani:2017bcs,Dimopoulos:2017ged}. 

The ultra slow-roll behaviour of the inflaton can also be understood from the shape of the potential after the last cliff, shown in Figure \ref{fig:zV}: by the end of the cliff, $\phi$ will be rolling very fast, allowing it to overshoot the local minimum, and pass through an inflection point -- shown in  Figure \ref{fig:zV} -- during which it decelerates with breaks on. As the acceleration term in the Klein Gordon equation dominates over the slope of the potential in this region, the velocity of the field decreases quickly\footnote{In single field inflation with canonical kinetic terms, the definition of the slow-roll parameter $\delta$ implies $\d \ln \dot{\phi} = -\delta~ \d N$. Assuming a constant $\delta$ during the ultra slow-roll regime gives $\epsilon \sim e^{-(2\delta)N}$.}, \ie $\epsilon \sim e^{-(2\delta)N}$ where $\delta \gtrsim 3$ (See Figure \ref{fig:eps1}). Note that although this regime is dubbed  `ultra slow-roll', the inflaton actually traverses the relevant part of the potential very quickly (in a few e-folds) by flying over its decreasing kinetic energy. After $\phi$ climbs the hill shown in Figure \ref{fig:zV}, the system is back into its slow-roll attractor regime and stays in it until inflation ends.

\medskip
In summary, the existence of sizeable non-perturbative corrections to the axion potential can lead to steep cliffs and gentle plateaus in the potential, including a local minimum, inflection point and maximum preceding the global minimum. We showed that in the presence of such a feature in the potential, the system enters into the ultra slow-roll regime for a few e-folds during which the slow-roll condition  $\delta \ll 1$ is violated. Such an order one violation of the slow-roll condition around a (near-)inflection point can lead to an enhancement of the primordial curvature power spectrum\footnote{See for example \cite{Motohashi:2017kbs} for a general discussion.}, in accordance with the arguments we reviewed in Section \ref{SecHeu}. We will discuss this enhancement for our model and the associated phenomenology for PBH as Dark Matter in Sections \ref{EPSPBH} and \ref{PBH}.  
Before focusing on this phenomenology at small scales, we need to make sure that the predictions of our model are in agreement with the observations at the CMB scales. This will be the topic of the following subsection.   

%

\subsection{Phenomenology at CMB scales}\label{cmb-sec}

We have seen that during the inflationary epoch, the slow-roll parameters undergo large oscillations when the field rolls down the steep cliffs and into the plateaus of the potential $\eqref{pot}$. However, during the short range of e-folds that is associated with CMB scales (\eg between $N=20$ and $N=30$ in both cases we considered), the slow-roll parameters can be small and smoothly varying \cite{Parameswaran:2016qqq}.  

We first estimate the CMB observables using the slow-roll approximation.  For this purpose,
we define the slow-roll parameters as
\beq\label{HJSR}
\epsilon = -\frac{\dot H}{H^2}, ~~~\delta = - \fr{\ddot{H}}{2 \,H\,\dot{H}},~~~\xi^2 = \fr{ \dddot{H} \dot{H} - \ddot{H}^2 }{2\,\dot{H}^2\,H^2},~~~\sigma^3 =  \fr{7
\dot{H} \ddot{H} \dddot{H}-2
\dot{H}^2 \ddddot{H}-5 \ddot{H}^3
}{4\,\dot{H}^3\,H^3}\,.
\eeq
In terms of these slow-roll parameters, the key observables describing the power spectrum of scalar and tensor perturbations are given by \cite{Stewart:1993bc,Easther:2006tv,Lyth:2009zz},
\begin{align}\label{obs}
\nn\Delta_s^2 &= \left[1 + (2-\ln 2-\gamma)(2\epsilon+\delta)-\epsilon\right]^2 \fr{H^2}{8\pi^2 \epsilon \Mp^2},\\
\nn\Delta_t^2 &=  \left[1 - (\ln 2+\gamma-1)\epsilon\right]^2 \fr{2}{\pi^2} \fr{H^2}{\Mp^2},~~~~ r = \fr{\Delta_t^2}{\Delta_s^2},\\
\nn n_s &= 1 -4\epsilon+2\delta - 2(1+\mathcal{C})\epsilon^2 - \fr{1}{2}(3-5\mathcal{C})\epsilon\delta +\fr{1}{2}(3-\mathcal{C})\xi^2,\\
\nn \alpha_s &= -2\xi^2 + 10\epsilon\delta -8\epsilon^2,\\
\nn \beta_s &= -32\epsilon^3 + 62\epsilon^2\delta -20\epsilon\delta^2+ 2\sigma^3-14\epsilon\xi^2+ 2\delta\xi^2,\\
\end{align}
where $\gamma$ is the Euler-Mascheroni constant and $\mathcal{C}\equiv 4(\ln 2 + \gamma)-5$. All time dependent quantities in the expressions above should be evaluated at the time of horizon crossing~\footnote{In single field inflation, we can always invert $\phi(N)$ to describe the slow-roll parameters in \eqref{HJSR} as a function of e-folds $N$.}  $N_{\rm hc}$ where the comoving pivot scale $k=k_*$ leaves the horizon $(a_{\rm hc} H_{\rm hc})^{-1}$. To determine the time of horizon crossing, we first use the numerical solutions to the FRW background equations \eqref{FRW} to describe the evolution of the Hubble slow-roll parameters in \eqref{HJSR} in terms of the number of e-folds $N$ and then impose the following measurements and bounds on the CMB observables from~\footnote{In the bumpy model we are investigating, the running of the running, $\beta_s$, is suppressed two orders of magnitude below $\alpha_s$. Therefore we will not consider the Planck results including $\beta_s$.} Planck 2015 (at $k_* = 0.05 ~{\rm Mpc}^{-1}  $ and $68\%~ {\rm CL}$ for TT+lowP+BAO)  \cite{Ade:2015lrj},
\bea\label{P15}
\nn \ln (10^{10}A_s) &=& 3.093\pm 0.034,\\
\nn n_s &=& 0.9673\pm 0.0043,\\
\nn \alpha_s &=& -0.0125 \pm 0.0091,\\
r &<& 0.166 \quad (\text{the Planck/KEK combined analysis gives }r<0.07), 
\eea
where $A_s$ is the amplitude of the curvature power spectrum. 

\begin{table}[t!]
\begin{center}
\begin{tabular}{ l c c }
\hline
\hline

 \cellcolor[gray]{0.9}  ~~&~~    \cellcolor[gray]{0.9} 
  Case 1: $\Mp/f =1.6$ ~~&~~   \cellcolor[gray]{0.9}  Case 2: $\Mp/f =1.7$\\

\hline
$N_*$ ~~&~~ $~~63.593$ ~~&~~ $54.015$ \\
$V_0^{1/4} $ ~~&~~~~~~~~~~ $9.98285\times 10^{-4}$ ~~&~~ $~~~~~~~~8.64799 \times 10^{-4}$ \\
$m $ ~~&~~ $~~2.563395477\times 10^{-6}$ ~~&~~ $~~~~2.12915358 \times 10^{-6}$ \\
$\Lambda_1$ ~~&~~~~~~~~~ $1.218164 \times 10^{-3}$ ~~&~~~~~~ $1.08856568 \times 10^{-3}$ \\
$\Lambda_2$ ~~&~~ $~~~~~8.9630812 \times 10^{-4}$ ~~&~~  $~~~~~7.1318712 \times 10^{-4}$\\
\hline
\hline
\end{tabular}
\caption{\label{tab:bparams} $N_*$ and the relevant mass scales in units of $\Mp$ in the axion potential \eqref{pot}.}
\end{center}			
\end{table}

In practice, we use the expression for scalar tilt in \eqref{obs} and impose the central value, $n_s = 0.9673$ to obtain the e-folding number at which the pivot scale crosses the horizon, $N_{\rm hc}$. This allows us to determine the number of e-folds before the end of inflation at which the pivot scale crosses the horizon, \ie $N_* \equiv N_{\rm tot} - N_{\rm hc}$. On the other hand, notice that in the previous section, we have scaled out the mass $m$ of the inflaton from all of our equations when we solve for the background evolution. Using the normalization of the power spectrum, $\Delta^2_s = 2.2 \times 10^{-9}$, we fix the mass scale $m$ which in turn allows us to determine $\Lambda_i$ from a given $\beta_i$. We list in Table \ref{tab:bparams} the values of $N_*$ and the parameters that we use in the axion potential \eqref{pot}. 

Calculating the observables by evaluating the expressions in \eqref{obs} at horizon crossing could lead to inaccurate results as the slow-roll parameter $\delta$ evolves considerably around this time (see Figures \ref{fig:eps}, \ref{fig:epsIS}) \cite{Leach:2001zf}. Therefore, to obtain reliable results, we have used the code~\footnote{Web page: $\mathsf{www.modecode.org}$}  $\mathsf{MultiModeCode}$ \cite{Mortonson:2010er,Easther:2011yq,Norena:2012rs,Easther:2013rva,Price:2014ufa,Price:2014xpa}. This program is  optimized for multi-field inflation, but its detailed implementation makes it easy to evolve single field background and perturbation equations at the linearized level with the bumpy potential defined in \eqref{pot}. Using the parameter sets given in Table \ref{tab:bparams}, we summarize the output of the program for the key observables in Table \ref{tab:bobs}. We see that all the observables associated with scalar fluctuations agree with the $2\sigma$ ($95 \%~ {\rm CL}$) limits of the Planck data (see \eg \eqref{P15}). The reason behind the large negative running can also be understood by evaluating the  slow-roll parameters \eqref{HJSR} at  horizon crossing:
\begin{table}[t!]
\begin{center}
\begin{tabular}{ l c c c}
\hline
\hline
 \cellcolor[gray]{0.9}
 Observables ~~&~~~  \cellcolor[gray]{0.9}   Case 1: $\Mp/f =1.6$ ~~&~~~  \cellcolor[gray]{0.9}  Case 2: $\Mp/f =1.7$\\

\hline
$~~~~~~n_s$ ~~&~~~~ $0.96717$ ~~&~~~~ $0.96408$ \\
$~~~~~~\alpha_s$ ~~&~~ $-0.03056$ ~~&~~ $-0.02987$ \\
$~~~~~~\beta_s$ ~~&~~ $~~~~2.78 \times 10^{-4}$ ~~&~~ $~~~~1.44\times 10^{-4}$ \\
$~~~~~~r$ ~~&~~  $~~~~7.88 \times 10^{-3}$ ~~&~~ $~~~~4.83 \times 10^{-3}$\\
$~~~~~~n_t$ ~~&~~  $~~-1.08 \times 10^{-3}$ ~~&~~ $~~-6.43 \times 10^{-4}$\\

\hline
\hline
\end{tabular}
\caption{\label{tab:bobs} Observables in bumpy axion inflation evaluated at the pivot scale $k_* =0.05~ {\rm Mpc^{-1}}$.}
\end{center}			
\end{table}
\beq
\nn \text{Case~1}:~~~~~ \epsilon \approx 5 \times 10^{-4},~~~\delta \approx -0.013,~~~\xi^2 \approx 0.004,~~~ 
\eeq 	
\beq
\text{Case~2}:~~~~~ \epsilon \approx 3 \times 10^{-4},~~~\delta \approx -0.027,~~~\xi^2 \approx 0.015~~.
\eeq 	
Notice that although the first slow-roll parameter $\epsilon$ is small, a large $\delta$ and $\xi^2$ leads to a large running $\alpha_s$ of the spectral index $n_s$. However, such large negative values of the running $\alpha_s$ are still within $2\sgm$ limits of the Planck data \eqref{P15}. Large values for the parameter $\alpha_s$ is a generic
prediction of our system based on a  potential with wiggles \cite{Kobayashi:2010pz, Parameswaran:2016qqq}.

For the two cases discussed  in this section, the field excursion during the observable range of inflation is $\Delta \phi> \Mp$. In particular, in Case 1  we  found that $\Delta \phi \simeq 8.7 \Mp$ whereas in Case 2, where inflation stops for a short time, $\Delta \phi \simeq 8.1 \Mp$.  Thus we have a smaller field range as compared to the smooth quadratic potential, which requires $\Delta \phi \simeq 15 \Mp$. The scale of inflation, $E_{\rm inf} \equiv (3H_*^2\Mp^2)^{1/4}$, on the other hand, is given by
\beq
\text{Case~1}:~~  E_{\rm inf} \simeq 4 \times 10^{-3} \Mp, \hskip1cm \text{Case~2}:~~ E_{\rm inf} \simeq 3.5 \times 10^{-3} \Mp.
\eeq
In the light of these results, it is interesting to note that the presence of sizeable non-perturbative corrections in the potential \eqref{pot} can give rise to a large field realization of axion inflation where a tensor-to-scalar ratio $r \approx 10^{-2}-10^{-3}$ can be obtained. In particular, we emphasize that although such values for the tensor-to-scalar ratio $r$ are typically small in comparison to the smooth monomial type potential $\phi^2$ \cite{Parameswaran:2016qqq}, they are within the sensitivity of next stage CMB experiments \cite{Abazajian:2016yjj}.

\subsection{Amplification of curvature perturbations at small scales}\label{EPSPBH}

In Section \ref{S:ultraslow}, we have seen that the background evolution of the inflaton in the bumpy potential \eqref{pot} can give rise to an ultra slow-roll era at small field values compared to the ones associated with the CMB. As we have emphasized before, this ultra slow-roll phase corresponds to a violation of slow-roll condition, where $\epsilon \ll 1$ but $\delta > 1$.
As we discussed in our initial Section \ref{SecHeu}, this regime can be characterised by an enhancement of the curvature
power spectrum, thanks to the contribution of the would-be decaying mode that is actually growing when the combination 
$z'/z = aH(1+\epsilon-\delta)$ is negative.  

\begin{figure}[t!]
\begin{center}
\includegraphics[scale=0.62]{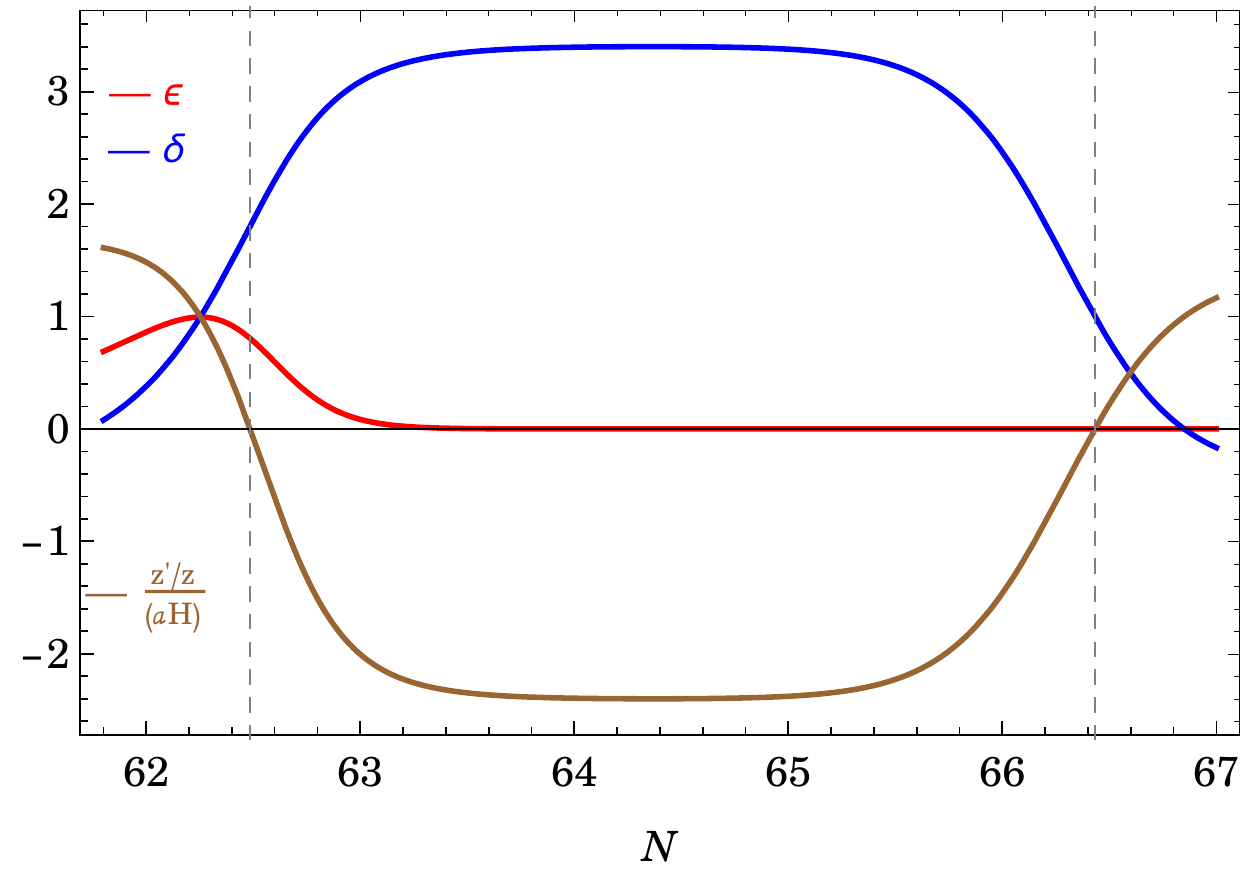}~~\includegraphics[scale=0.62]{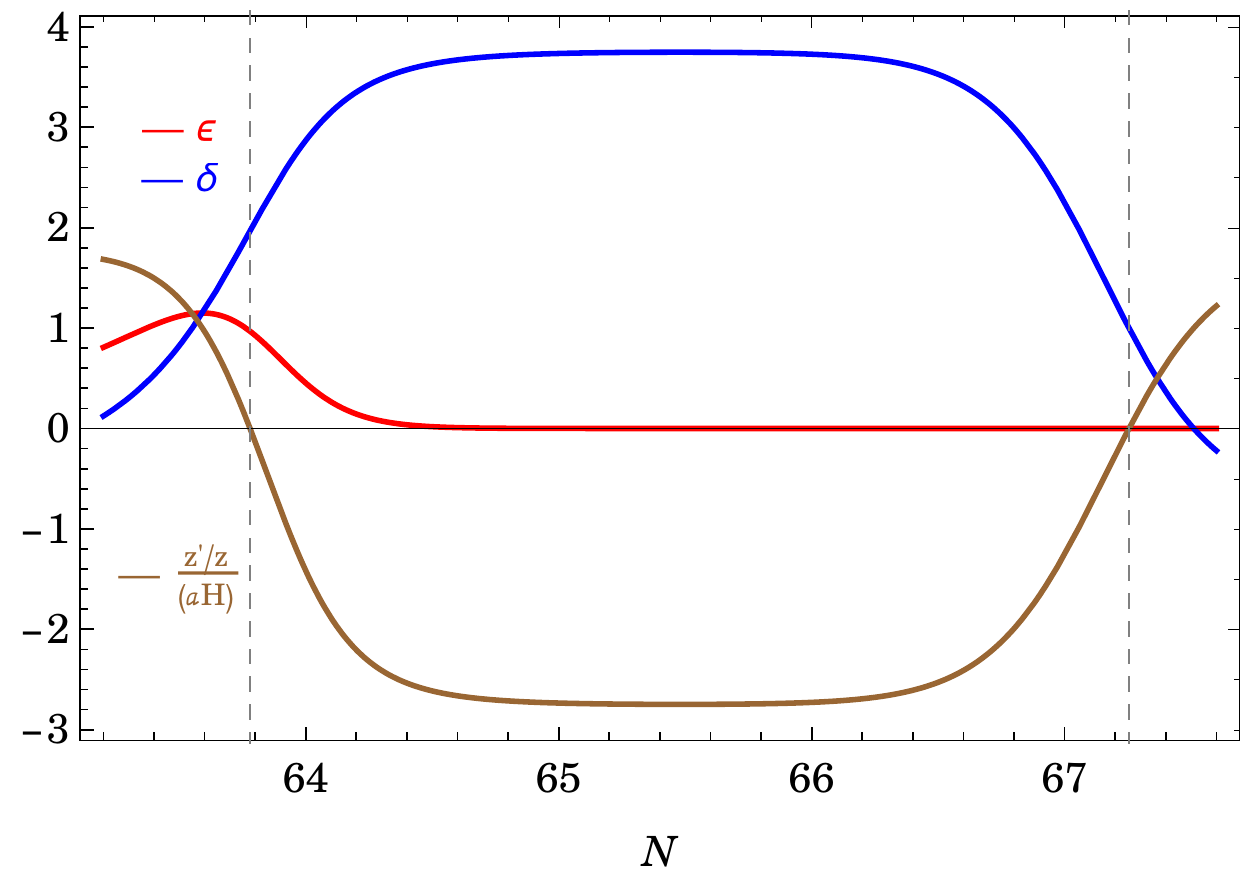}
\end{center}
\caption{Evolution of the $1+\epsilon-\delta$ as a function of e-folds around the ultra slow-roll regime for the examples we presented in Section \ref{BBack}. $z'/z < 0$  between the two vertical dashed lines.\label{fig:zp}}
\end{figure}

Both of the examples we presented in Section \ref{BBack} -- in eqs \eqref{cca1} and \eqref{cca2} --  fit well 
 with the argument developed in Section \ref{SecHeu}, during  the phases
  where the system first enters  a fast-roll regime ($\delta = 1$), followed by an ultra slow-roll era, $\delta\gtrsim 3$. To illustrate this fact, in Figure \ref{fig:zp}, we superimpose the plot of the function $(aH)^{-1}~z'/z$ with that of $\epsilon$ and $\delta$ for the two representative cases we discussed  in Section \ref{BBack} (see \eg Figures \ref{fig:eps1} and \ref{fig:eps1IS}). We observe that a turn around in $|z|$ occurs shortly after the time when $\delta >1$ and $z'/z$ continues to be negative into the ultra slow-roll regime where $\delta \gtrsim 3$, implying a growth in $\mathcal{R}_k$ until the system returns back to the slow-roll phase.
Notice also that the shape of the $z'/z$ in this region is  identical to the one of $\delta$, which implies that the negativity of this function is mainly dictated by a large value of  $\delta$. 

\begin{figure}[t!]
\begin{center}
\includegraphics[scale=0.9]{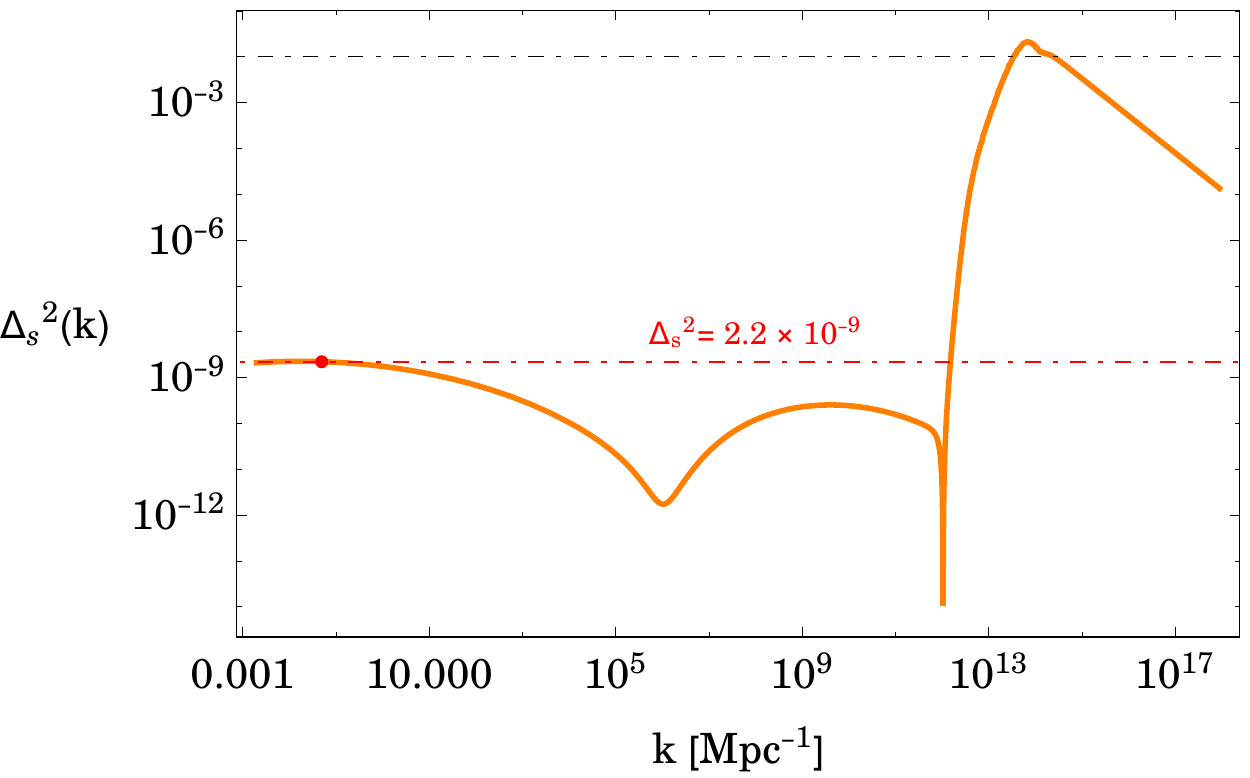}
\includegraphics[scale=0.9]{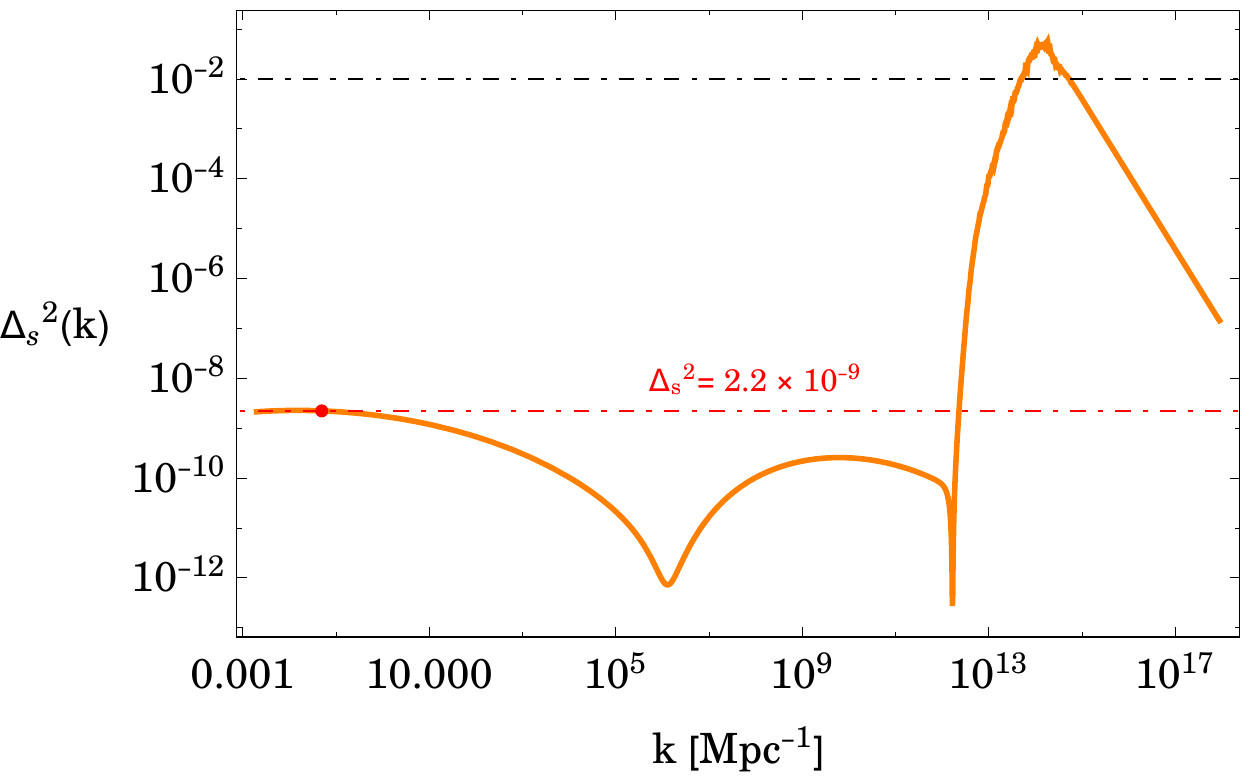}
\end{center}
\caption{Power spectrum of scalar curvature perturbation in bumpy axion inflation with the parameter choices shown in Table \ref{tab:bparams} (Case 1-top panel, Case 2-bottom panel). Red dot in the graph represent the point where $\Delta_s^2 = 2.2 \times 10^{-9}$ and  $k_*  = 0.05 ~{\rm Mpc^{-1}}$. \label{fig:ps}}
\end{figure}

In addition to the heuristic understanding we presented above, we also analysed numerically the full power spectrum of curvature perturbations defined by $\Delta^2_s (k) \equiv k^3 |\mathcal{R}_k|^2/(2\pi^2)$. Given the complexity of the scalar potential and of the background dynamics we described above, we solve the linearized perturbation equations numerically using the $\mathsf{MultiModeCode}$ to obtain the full power spectrum. The resulting scalar power spectra, using the parameters provided in Table \ref{tab:bparams}, is shown in Figure \ref{fig:ps}. We see that both of the scalar power spectra start to grow at comoving scales $k> 10^{13}~ {\rm Mpc^{-1}}$ and peak around $k \approx 7 \times 10^{13} - 10^{14} ~{\rm Mpc^{-1}}$, reaching  a value of $\Delta_s^2\approx 10^{-2}$. As we have explained, the reason for the growth in the power spectrum is the ultra slow-roll regime while $\phi$ overshoots a local minimum of the bumpy potential at Planckian field values (see \eg Figure \ref{fig:zV}). In the following section, we  investigate the phenomenology arising from these peaks in the power spectrum.

\subsection{PBHs from non-perturbative effects in axion inflation}\label{PBH}

PBHs may have formed in the very early Universe if a sufficiently large amplitude of primordial fluctuation is generated at small scales ($k\gg k_*= 0.05~ {\rm Mpc^{-1}}$) during inflation. In an inflationary Universe such a mode is stretched outside the comoving horizon and it re-enters the horizon at a later time after the end of inflation. If the amplitude of these fluctuations is significant, there will be regions in the Universe where the density of matter is so large that it can collapse to form PBHs upon horizon re-entry \cite{Hawking:1971ei,Carr:1974nx}. PBHs can have observational implications at the current epoch by contributing to the present ``cold'' Dark Matter density if they are massive enough to avoid Hawking evaporation \cite{Carr:1975qj,Ivanov:1994pa}.

In the simplest case, the mass of the resulting PBHs is assumed to be proportional to mass inside the Hubble volume at the time of horizon re-entry (at the time of PBH formation) of a mode with wavenumber $k$: 
\beq\label{mk}
M(k) = \gamma \fr{4\pi}{3}\rho H^{-3} \bigg\rvert_{k=a_{f} H_{f}} = \gamma~ M_{\rm eq}^{\rm H} \left(\fr{\rho_{f}}{\rho_{\rm eq}}\right)^{1/2} \fr{H_{\rm eq}^2}{H_{f}^2}
\eeq
where $M_{\rm eq}^{\rm H}$ is the horizon mass at the time of matter-radiation equality and the subscripts ``$f$'' and ``${\rm eq}$'' denote quantities evaluated at the time of PBH formation and matter-radiation equality, respectively.   
 Using the conservation of entropy, $g_s(T)~ T^3~ a^3 =const.$ and the scaling of the energy density with the temperature in the radiation dominated era, $\rho \propto g_*(T)~ T^4$  , the mass of the PBHs can be expressed in terms of the comoving wavenumber $k$ as 
\begin{align}
\nn M(k) &= \gamma ~M_{\rm eq}^{\rm H} \left(\fr{g_{*}(T_f)}{g_{*}(T_{\rm eq})}\right)^{1/2} \left(\fr{g_{s}(T_{\rm eq})}{g_{s}(T_{f})}\right)^{2/3} \left(\fr{k_{\rm eq}}{k}\right)^{2}\\ 
&\simeq  1.6 \times 10^{18}~{\rm g}~ \left(\fr{\gamma}{0.2}\right) \left(\fr{g_{*}(T_f)}{106.75}\right)^{-1/6}\left(\fr{k}{5.5\times 10^{13}~ {\rm Mpc^{-1}}}\right)^{-2},
\end{align}
where in the second line we assumed $g_*(T) = g_s(T)$ and used $M_{\rm eq}^{\rm H} \simeq 5.6 \times 10^{50} {\rm g}$ \cite{Nakama:2016gzw}, $g_*(T_{\rm eq}) = 3.38$, $k_{\rm eq} = 0.07~\Omega_m h^2~ {\rm Mpc^{-1}}$. The value of the constant of proportionality  $\gamma = 0.2$ is suggested by the analytical model in \cite{Carr:1975qj} for PBHs formed during the radiation dominated era. 

The standard treatment of PBH formation is based on the Press-Schechter model of the gravitational collapse that is used widely in large-scale structure studies \cite{Press:1973iz}. In this context, the energy density fraction in PBHs of mass $M$ at the time of formation, which is denoted by $\beta(M)$, is given by the probability that the fractional overdensity $\delta \equiv \delta \rho/\rho$ is above a certain threshold $\delta_c$ for PBH formation. For Gaussian primordial fluctuations\footnote{The presence of local non-Gaussianity can significantly alter the PBH abundance \cite{PinaAvelino:2005rm,Young:2013oia, Tada:2015noa, Young:2015kda, Young:2015cyn, Franciolini:2018vbk}.  Although non-Gaussianities at CMB scales for our model are small ($|f_{\rm NL}| \sim 10^{-4}$ at $k_* = 0.05~{\rm Mpc^{-1}}$), consistently with Maldacena's consistency relation for single-field slow-roll models \cite{Maldacena:2002vr}, non-Gaussianities during the ultra slow-roll regime at small scales may be sizeable \cite{Namjoo:2012aa}.  However, if the transition between ultra slow-roll and slow-roll is smooth, as is the case here, then those non-Gaussianities are washed out by subsequent evolution \cite{Cai:2017bxr}, and thus we can neglect the effects of non-Gaussianities when calculating PBH abundances.  See also \cite{Bravo:2017gct} for a discussion on the vanishing of observable primordial local non-Gaussianity in canonical single-field inflation.}, $\beta(M)$ is  given by
\bea\label{beta}
\nn\beta(M(k))\equiv \fr{\rho_{\rm PBH}}{\rho}&=&2 \int_{\delta_c}^\infty \fr{\d \delta}{\sqrt{2\pi}\sigma(M(k))}\exp\left(-\fr{\delta^2}{2\sigma^2(M(k))}\right),\\
&= & \sqrt{\fr{2}{\pi}}\fr{\sigma(M(k))}{\delta_c} \exp\left(-\fr{\delta_c^2}{2\sigma^2(M(k))}\right)
\eea
where the factor of 2 accounts for locally under threshold regions collapsing in globally over threshold regions and we have assumed $\delta_c > \sigma$ in the second line of \eqref{beta}. The value of the $\beta(M)$ is uniquely determined by the variance $\sigma^2(M(k))$ which is assumed to be coarse-grained variance smoothed on a scale of $R = k^{-1}$. During the radiation dominated era, it is given by the following expression \cite{Young:2014ana},
\beq\label{var}
\sigma^2(M(k))= \fr{16}{81}\int_{0}^{\infty} \d\ln q~\left(\fr{q}{k}\right)^4 \Delta_s^2(q)~ W(q/k)^2,
\eeq
where $\Delta^{2}_s$ is the power spectrum of curvature perturbation and $W(x)$ is a smoothing window function which is usually taken to be of the Gaussian form, $W(x) = \exp(-x^2 /2)$.

At the time of their formation, a fraction, $\gamma \beta(M(k))\rho |_{k=a_f H_f}$ , of the total energy in the Universe turns into PBHs. After their formation, $\beta$ grows inversely proportional to the cosmic tempertaure ($\propto a$) until matter-radiation equality,  since PBHs essentially behave as pressureless dust ($\rho_{\rm PBH}\propto a^{-3}$). Therefore, the fraction of PBH abundance in Dark Matter today can be determined by a simple red-shifting relation \cite{Inomata:2017okj}
\bea\label{f}
\nn\fr{\Omega_{\rm PBH}(M(k))}{\Omega_{\rm DM}}&=&\left(\fr{T_f}{T_{\rm eq}}\fr{\Omega_m}{\Omega_{\rm DM}}\right)\gamma \beta(M(k)),\\
&\simeq &\left(\fr{\beta(M(k))}{10^{-15}}\right)~ \left(\fr{\gamma}{0.2}\right)^{3/2} \left(\fr{g_{*}(T_f)}{106.75}\right)^{-1/4} \left(\fr{M(k)}{1.6\times 10^{18}~ {\rm g}}\right)^{-1/2},
\eea
where $T_f$ is the temperature of the plasma at the time of PBH formation and $T_{\rm eq}$ is the temperature at matter-radiation equality. In order to determine the ratio of the total energy density in PBHs today to that of Dark Matter, we  integrate over all masses $M$,
\beq
\fr{\Omega_{\rm PBH}^{\rm tot}}{\Omega_{\rm DM}} = \int \d\ln( M(k))~ \fr{\Omega_{\rm PBH}(M(k))}{\Omega_{\rm DM}}.
\eeq

It is clear from the expression in \eqref{f} that PBHs of mass $M \sim 10^{18}~ {\rm g}$ can constitute a significant fraction of Dark Matter density today if $\beta(M)$ is within a couple of orders of magnitude of $\sim 10^{-15}$. On the other hand, it is worth emphasizing that the PBH abundance is exponentially sensitive to the critical threshold density for collapse $\delta_c$ and the variance $\sigma^2$ (see  equation \eqref{beta}). In the following, to estimate the total PBH abundance with respect to Dark Matter abundance today, we will take values of $\delta_c$ within the range $\delta_c = 0.3 -0.5$ as suggested in \cite{Carr:1975qj,Harada:2013epa,Musco:2012au}. For these values of $\delta_c$, one requires $\sigma^2(M)\sim 10^{-2} - 10^{-3}$ to reach  the required level $\beta(M)$ on the relevant scale $k$ (or $M$). This in turn arises from a power spectrum in equation \eqref{var} that is also of the order of $10^{-2}$. 

\begin{figure}[t!]
\begin{center}
\includegraphics[scale=0.63]{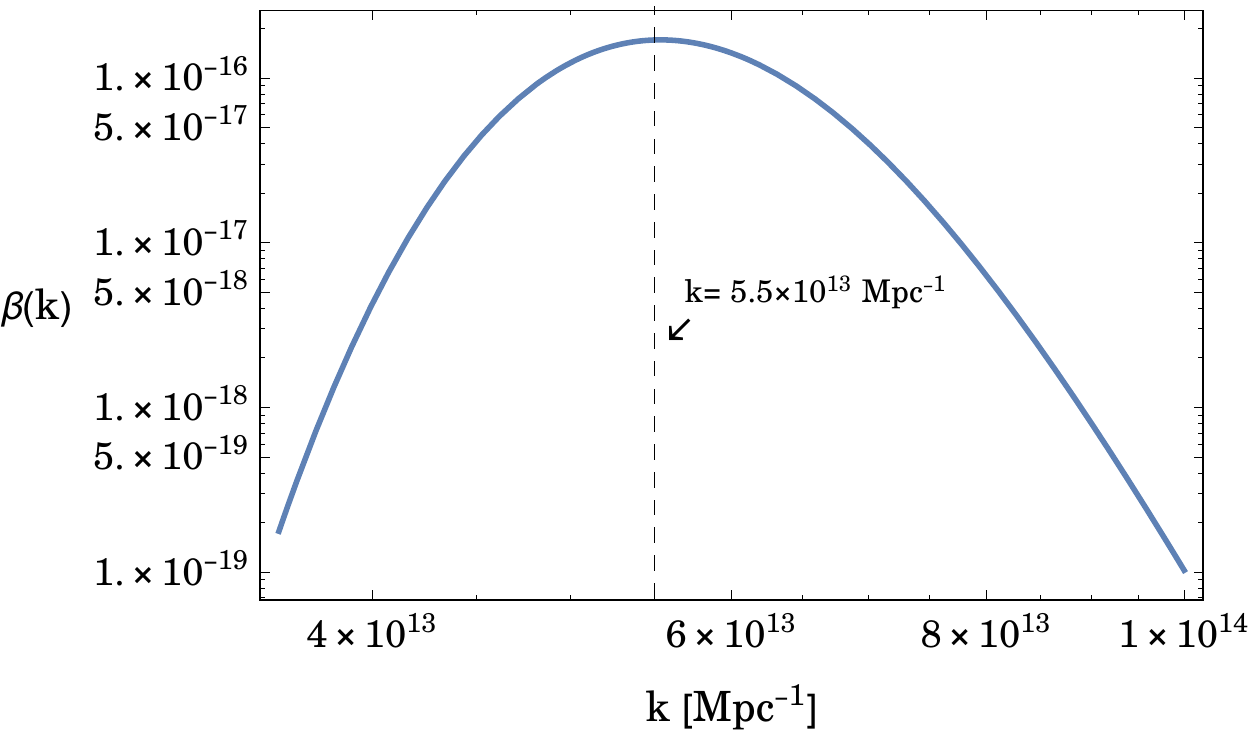}
\includegraphics[scale=0.63]{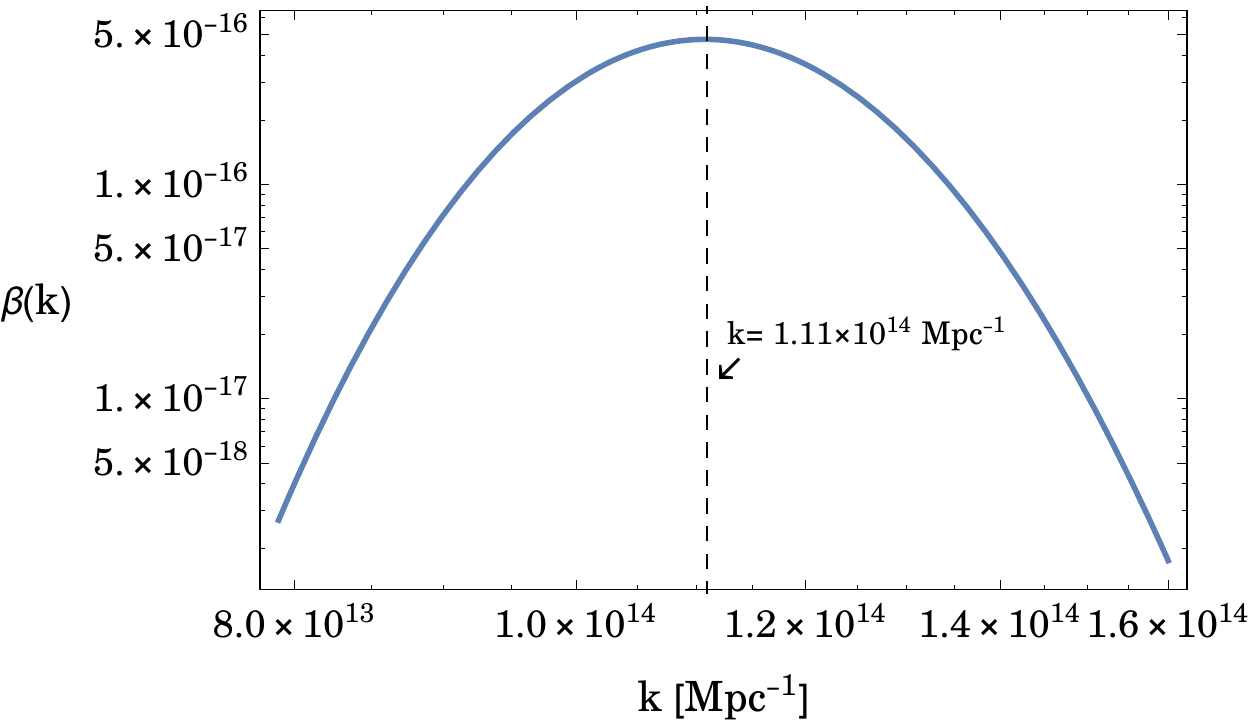}
\end{center}
\caption{$\beta$ as a function of the smoothing scale $k$. The values of $k$ where $\beta(k)$ has a peak is also shown with dashed vertical lines. \label{fig:beta}}
\end{figure}

Figure \ref{fig:ps}  shows that  both  examples we presented in the previous sections satisfy this criterion where both power spectra have sharp peaks around $k \sim 5\times 10^{13}- 10^{14}~{\rm Mpc^{-1}}$. To illustrate this further, we calculate $\beta(k)$ in \eqref{beta} by numerically integrating the variance in \eqref{var} using the full power spectrum $\Delta^{2}_s(q)$. The resulting  $\beta(k)$ for a range of comoving wave numbers including its peak is shown in Figure \ref{fig:beta}. Only the range of $k$ values shown in this plot have a significant contribution to the total PBH abundance which can be calculated numerically using the fact that $\Omega_{\rm PBH}/\Omega_{\rm DM} \propto \int \d k~ \beta(k)$.   
In Table \ref{tab:apbh}, we summarize these results on the fraction of PBHs in Dark Matter today and the peak value for the mass of PBHs (using \eqref{mk} with $M_{\rm peak}\equiv M(k=k_{\rm peak})$) obtained from the inflationary models we considered in Table \ref{tab:bparams}. 
\begin{table}[t!]
\begin{center}
\begin{tabular}{ l c c c}
\hline
\hline
 \cellcolor[gray]{0.9}
Cases ~~&~~   \cellcolor[gray]{0.9} $\delta_c$  ~~&~~  
 \cellcolor[gray]{0.9}
$M_{\rm peak}/M_\odot$ ~~&~~
 \cellcolor[gray]{0.9}  $\Omega_{\rm PBH}^{\rm tot}/\Omega_{\rm DM}$\\

\hline
$\Mp/f =1.6$ ~~&~~ $~~0.34$ ~~&~~ $~~8 \times 10^{-16}$ ~~&~~ $0.113$ \\
$\Mp/f =1.7$ ~~&~~ $~~~0.5$ ~~&~~ $~2 \times 10^{-16}$~~&~~ $0.514$ \\

\hline
\hline
\end{tabular}
\caption{\label{tab:apbh} The two different choices of the critical threshold overdensity $\delta_c$ and the corresponding total abundance of PBHs for the models considered in Table \ref{tab:bparams}. The peak value of the mass of the relevant PBHs which are obtained from the equation \eqref{mk}, is also shown. }
\end{center}			
\end{table} 

\subsubsection{Observational constraints on PBH abundance}

 We have seen that  axion inflation with subleading non-perturbative corrections can give rise to PBHs of mass $M 
\simeq 3.9 \times 10^{17} - 10^{18} g$, which  can constitute an $\mathcal{O}(1)$ fraction\footnote{Recall that $\Omega_{\rm PBH}^{\rm tot}/\Omega_{\rm DM}$ is exponentially sensitive to the value of $\delta_c$ which can be adjusted within the suggested range in the literature to increase PBH abundance. } of Dark Matter today. It has been pointed out that such compact objects can induce features in the photon spectrum of the gamma-ray bursts that occur at cosmological distances \cite{gould1992femtolensing}. The angular separation of these photon sources which would be lensed by such small PBHs is around the femto scale, hence the class of constraints obtained by these sources is
 called ``femto-lensing''. Recent analysis on femto-lensing of gamma ray bursts shows that PBHs in the mass range $5 \times 10^{17} - 10^{20}~{\rm g}$ cannot constitute more than $10 \%$ of Dark Matter \cite{Barnacka:2012bm}. In particular, for the realization of axion inflation with the first parameter set given in Table \ref{tab:bparams}, the PBH abundance cannot be  much higher than the level shown in Table  \ref{tab:apbh}. However, the constraints\footnote{Note however that contraints  on the PBH fraction tend to become stronger for non-monochromatic mass functions in this range of PBH masses \cite{Carr:2017jsz}.} become weaker for the smaller mass PBHs found in the case where inflation terminates for a short period of time, and thus do not exclude the interesting possibility that these tiny PBHs provide a significant fraction of Dark Matter (see \eg Figure 1 of \cite{Ballesteros:2017fsr} or Figure 4 of \cite{Carr:2016drx}). 

In summary, we have found that axion inflation with subleading but significant non-perturbative corrections is capable of generating a population of light PBHs ($M \simeq 10^{-15}-10^{-16} M_\odot$) that might account for a considerable fraction of Dark Matter in the Universe. It is also interesting to note that there is no known astrophysical mechanism that can produce black holes with
   such small mass.  Here we have proposed a string theory inspired primordial mechanism that can produce small mass black holes.

\subsubsection{Implications for reheating}
In the bumpy axion inflation we are considering here,
in order to both account for a large enough amplification in the scalar power spectrum at small scales and an agreement with observations at CMB scales by Planck, the observable scales associated with the CMB had to leave the horizon at values of\footnote{In fact, in order to increase the amplification of power spectrum on small scales, one is required to tune (by increasing) $\Lambda_1$ at the level shown in Table \ref{tab:bparams} which in turn increases the total number of e-folds during inflation without altering $N_{\rm hc}$, implying a larger $N_* = N_{\rm tot} - N_{\rm hc} $.} $N_*=N_{\rm tot} - N_{\rm hc}$ given in Table \ref{tab:bparams}. In this section, we will discuss the theoretical implications of these values on the reheating phase after inflation.

In general, there is a theoretical uncertainty in determining $N_*$ due to the unknown thermal history of the Universe after inflation. Parametrizing our ignorance about the post-inflationary Universe by an average equation of state $w_p$ and the energy density at the time of reheating $\rho_{\rm rh}$, we can quantify this uncertainty by using the matching equation \cite{Liddle:2003as,Adshead:2010mc,Easther:2011yq},
\beq\label{Match}
N(k) = -71.21 -\fr{1}{4}\ln\left(\fr{ \rho_{\rm end}}{\Mp^4}\right)-\ln\left(\fr{k}{H_*}\right)+ \fr{(1-3w_p)}{12(1+w_p)}\ln\left(\fr{\rho_{\rm rh}}{ \rho_{\rm end}}\right).
\eeq 
Denoting by $N^{\rm inst}_*$ the value of $N_*$ in the case of instantaneous reheating, \ie $\rho_{\rm rh} \to \rho_{\rm end}$, we can write:
\beq
\delta N \equiv N_* - N^{\rm inst}_* = \fr{(1-3w_p)}{12(1+w_p)}\ln\left(\fr{\rho_{\rm rh}}{ \rho_{\rm end}}\right).
\eeq
On the other hand, earlier in our discussion we have assumed that the PBH formation occurs during the radiation dominated era after thermalization is complete. This implies that the temperature of the plasma at the time of PBH formation is less than the temperature at the time of reheating, $T_{\rm rh}\geq T_f$, where we can quantify $T_f$ using the general expression in \eqref{mk} as 
\beq\label{TF}
T_f \simeq 2.65 \times 10^{6}~ {\rm GeV} \left(\fr{\gamma}{0.2}\right)^{1/2} \left(\fr{g_{*}(T_f)}{106.75}\right)^{-1/4} \left(\fr{M}{8 \times 10^{-16}~ M_\odot}\right)^{-1/2}.
\eeq
With the choices of parameters shown in Table \ref{tab:bparams}, we numerically obtained the energy density at the end of inflation $\rho_{\rm end}\equiv 3/2~ V_{\rm end}$ as
\beq
\text{Case 1}:~ \rho_{\rm end}= 2.7\times 10^{-12}~\Mp^4,~~~~~~~\text{Case 2}:~ \rho_{\rm end}= 1.6\times 10^{-12}~\Mp^4.
\eeq    
Therefore for both cases, using the expression for $T_f$ with the fiducial choices of parameters in \eqref{TF}, $\delta N$ can be estimated as 
\beq\label{dN}
\delta N \simeq \fr{(1-3w_p)}{12(1+w_p)} \ln \left(10^{-35}\left(\fr{g_{*}(T_{\rm rh})}{106.75}\right)\left(\fr{T_{\rm rh}}{T_f}\right)^{4}\right).
\eeq
On the other hand, using our numerical results of Section \ref{BBack}, in the bumpy axion inflation we obtained $N^{\rm inst}_* \simeq 57$ at $k_* = 0.05~{\rm Mpc^{-1}}$, corresponding to $\delta N \simeq 6.59$ (Case 1) and $\delta N \simeq -3$ (Case 2) by reading the values of $N_*$ from Table \ref{tab:bparams}.

Note from the equation \eqref{dN} that to obtain a positive $\delta N$, a non-trivial average equation of state is required in the post-inflationary\footnote{A non-standard post-inflationary evolution can give rise to further changes in $\delta N$ as discussed in \cite{MZ} for a string theory motivated scalar-tensor evolution after inflation.} Universe, \ie $1/3 < w_p < 1$ since $\rho_{\rm rh}\leq \rho_{\rm end}$. Assuming the smallest available value of the argument of the logarithm (\ie $T_{\rm rh} = T_f$) in \eqref{dN}, one can reach to a maximum value of $\delta N\simeq 6.7$ by an avarage equation of state correponding to a stiff fluid $w_p =1$. This implies that interesting phenomenology in the bumpy axion inflation with the parameter choices in Case 1 is only possible if a kination type of fluid \cite{Joyce:1996cp,Ferreira:1997hj} dominates the energy density of the post-inflationary Universe until $T_{\rm rh}\simeq  10^{6}~{\rm GeV}$. In the Case 2 however, theoretical requirements on the post-inflationary evolution is much less restrictive because one can easily accomodate $\delta N \simeq - 3$ for a wide range of reheating temperatures $T_{\rm rh}$ assuming a non-standard cosmology\footnote{Such cosmologies could be favoured by certain string theory constructions, see for example \cite{Easther:2013nga} and other possible observational effects that might arise in these scenarios \cite{Erickcek:2011us,Fan:2014zua,Erickcek:2015jza}.} with an average equation of state satisfying $0\leq w_p<1/3$.\\

\section{DBI inflation with steps in the warp factor}

In the previous section, we have seen that suitable choices of  string inspired scalar potentials can lead
to the production of PBHs,  through a  small scale  enhancement of the  curvature power spectrum induced
by rapid changes in some of the slow-roll parameters. String
theory also motivates models of inflation with non-standard kinetic terms, with a scalar Lagrangian  expressed as 
 \beq \label{non-lag}
{\cal L}_\phi = \sqrt{-g} \,P(X,\phi)
\eeq
where $X\equiv \frac12 (\partial\phi)^2$, and $P$ a certain function $P$ of $X,\,\phi$. 
In this section, we begin to explore
whether appropriate choices of the kinetic function  $P$ can provide the kind of violations of slow-roll conditions which  enhances  the scalar
power spectrum, motivated by
a generalization of 
 the argument  presented in Section \ref{SecHeu}. Our analysis here will be  only qualitative -- we do not numerically compute the power spectrum in this case -- but serves
as starting point for further  quantitative studies of black hole production in scalar-tensor theories of single field inflation with non-standard kinetic terms.

\subsection{Background and perturbations with non-canonical kinetic terms}
We start by writing 
the homogeneous equations of motion associated with the general Lagrangian with non-canonical kinetic terms, eq (\ref{non-lag}), minimally coupled with Einstein gravity:
\bea
H^2= \frac{8\pi G}{3} \rho, \\
\dot H = -4\pi G (\rho + P), \\
 \dot\rho = -3H(\rho + P) \,.
\eea
The energy density is  
\beq\label{rhox}
\rho = 2X P_{,X} - P \,,
\eeq
while the function $P$ in eq \eqref{non-lag} plays the role of pressure. The dot indicates derivatives along physical time $t$. 
This system is characterized by a sound speed\footnote{The implications  of a smaller than unity speed of sound for the cosmological observables ($n_s, r, \alpha_s$) in a model independent large-$N$ approach was studied in \cite{IZ}, while
 frameworks to study large deviations
from a slow-roll regime in  similar contexts were developed in  \cite{RenauxPetel:2008gi,Emery:2012sm,Emery:2013yua}.}
 $c_s$ defined by 
\beq
c_s^2 = \frac{P_X}{\rho_X} = \frac{P_X}{P_X +2X P_{XX}}\,.
\eeq

Following the formalism developed by Garriga and Mukhanov \cite{GM}, the equation for the curvature perturbation $\mathcal{R}$  generalises eq.~\eqref{mez1} and reads in this case
 \beq
 \label{mez2}
\mathcal{R}''_k + 2\fr{z'}{z}
\mathcal{R}'_k + c_s^2\,k^2
\mathcal{R}_k = 0, 
 \eeq
where now
\beq
\fr{z'}{z}= aH \left(1+\epsilon-\delta -s\right) \label{defzpz2}\,,
\eeq
with the prime corresponding to derivatives along conformal time $\tau$. 
The slow-roll parameters are defined as
\bea
 \epsilon = -\frac{\dot H}{H^2} \,, \qquad 
 \delta = - \frac{\ddot H}{2 H \dot H} \,,\qquad 
  s = \frac{\dot c_s}{H c_s} \,.
\eea

At this point, we  can generalise the  arguments  we introduced in  Section \ref{SecHeu}, where we 
have seen  that the spectrum of curvature fluctuations can be   enhanced   at  small  
scales by violating the slow-roll conditions and
  changing the sign in the quantity $z'/z$ (see the discussion around eq.~\eqref{condtste1}). In the 
case of non-canonical kinetic terms, we have one additional quantity to use --  the parameter $s$ which can turn large -- and hence  novel possibilities for producing PBHs.
A full analysis of the phenomenological consequences of models based on this approach goes outside the scope of this paper:
 we limit ourselves to  illustrating the effect of a varying speed of sound in a representative example.

\subsection{Enhancement of  curvature fluctuations in DBI inflation } 

The most famous example of inflation with non-standard kinetic terms motivated by  string theory is Dirac-Born-Infeld (DBI) inflation \cite{ST,AST}. In this scenario, a probe D3-brane moves in the warped throat of a flux compactification in type IIB string theory. The dynamics is described by the DBI and Wess-Zumino actions, which give rise to:
\beq
P(X,\phi) = \frac{1}{h(\phi)}(1-\gamma^{-1}) -V(\phi)\,,
\eeq 
 where $h(\phi)$ is the warp factor, which depends only on $\phi$, and $\gamma$ is defined as
 \beq
 \gamma^{-2} = 1+2Xh(\phi) \,.
 \eeq
 In this set-up, the speed of sound $c_s^2 = \gamma^{-2}$. 
 Using \eqref{rhox}, we  find that the energy  density is given by:
\beq
\rho = \frac{1}{h(\phi)}(\gamma-1) + V(\phi)\,.
\eeq

We now consider what realisations of DBI setups can exhibit a transiently large slow-roll parameter $s$, which as we have observed can change the sign of the quantity $z'/z$ \eqref{defzpz2} in the eq \eqref{mez2} governing the curvature perturbation, driving a growth in the latter.  A large value for $s$ can be induced by systems where the warp factor experienced by the moving
D-brane has features 
 \cite{Bean:2008na,MHA}.    In particular, given that $c_s = 1/\gamma = \sqrt{1 - h(\phi)\dot{\phi}^2}$, a large $s=\dot {c_s}/(c_s H)$ could be achieved if $h(\phi)$ has a sharp decline.  Such a feature in the warp factor could arise, for instance, if the D-brane that drives inflation travels down a double warped throat, sourced by two separated stacks of D-branes/localised fluxes \cite{Franco:2005fd, Cascales:2005rj}.  Another possibility is that during the DBI-brane's journey through the warped throat, some D-branes or flux at the bottom of the throat annihilate with some $\overline{\text{D}}$-branes, either perturbatively via D-brane--$\overline{\text{D}}$-brane annihilation, or non-perturbatively via the KPV instability \cite{Kachru:2002gs}, thus reducing the strength of the warp factor.

\begin{figure}[t!]
\begin{center}
\includegraphics[scale=0.72]{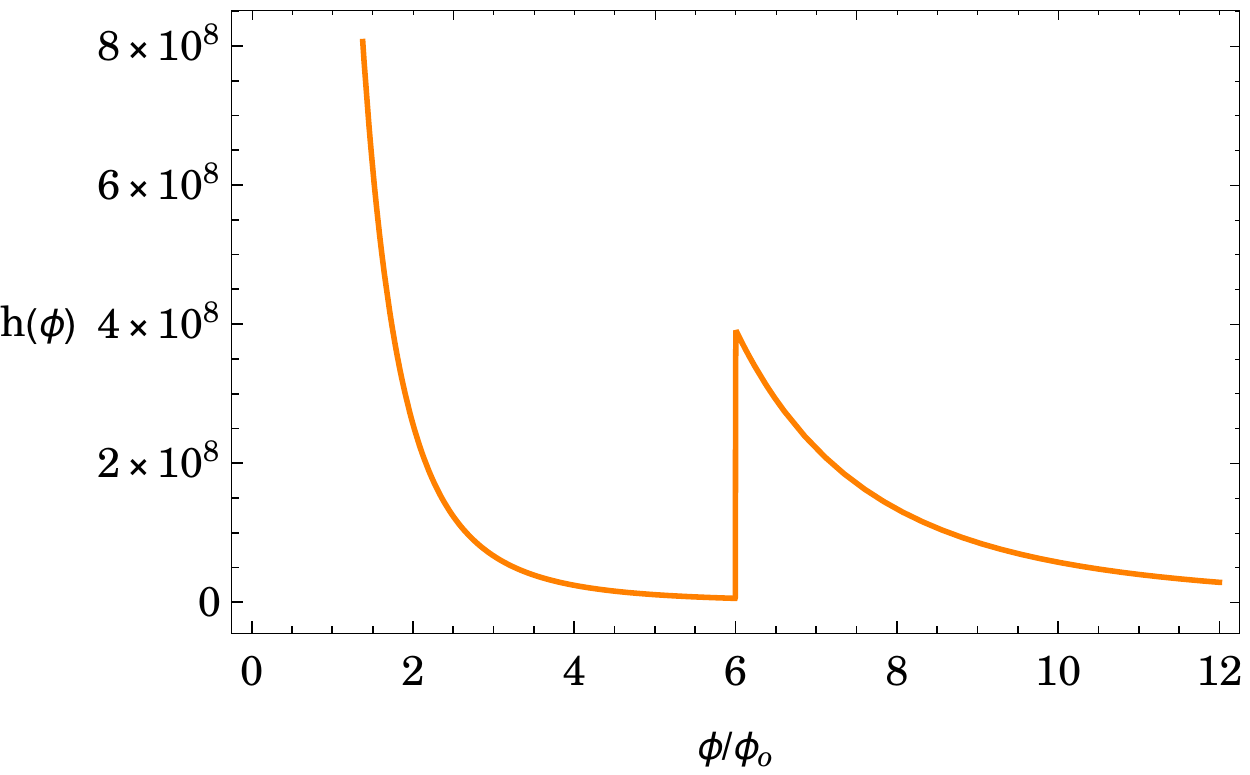}
\end{center}
\caption{Warp factor $h(\phi)$ \eqref{WF} as a function of $\phi/\phi_{_0}$ for the following parameter choices $\lambda=10^{10}$, $a=70$, $b=69$, $c=2000$, $d=6$.
 }\label{fig:WF}
\end{figure}

	We  model these features by adding a step into an adS warp factor (though it would be important to understand more accurately how to realistically describe  transitions in the warped geometry):
\beq\label{WF}
h(\phi) = \frac{a \lambda}{(\phi/\phi_{_0} +1/2)^4} -\frac{b \lambda}{(\phi/\phi_{_0} +1/2)^4}  
\left( 1-\frac{1}{1+e^{-c(\phi/\phi_{_0} -d)}} \right)\,.
\eeq
This can be seen as a finite tip version of an adS warp factor with a step feature. The parameters $a,\,b,\,c,\,d,\lambda$, and $\phi_{_0}$ are constants expressed
in appropriate units.  
We plot  this warp factor in Figure \ref{fig:WF}.
For simplicity, we will further assume a quadratic scalar potential:
\bea
V(\phi) = \frac{1}{2} m^2\phi^2\,.
\eea

\begin{figure}[h!]
\begin{center}
\includegraphics[scale=0.59]{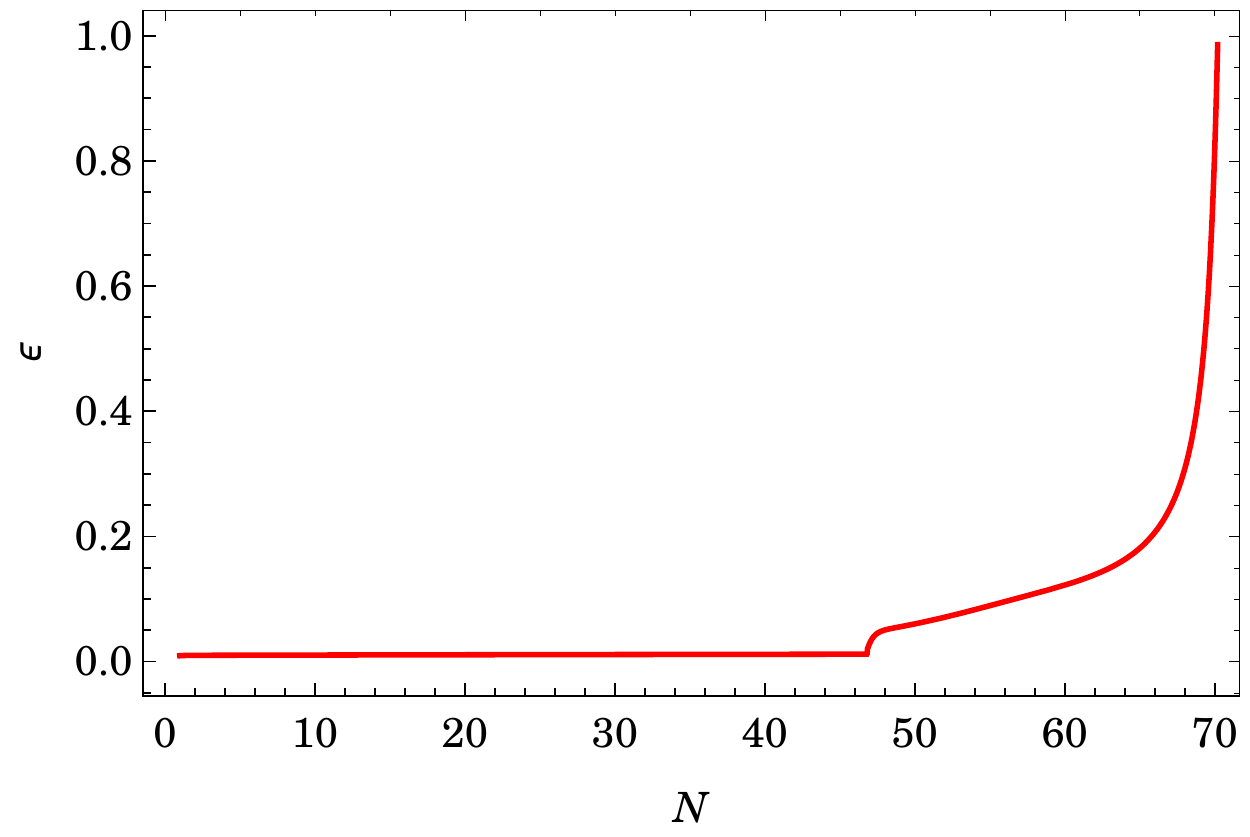}  \includegraphics[scale=0.59]{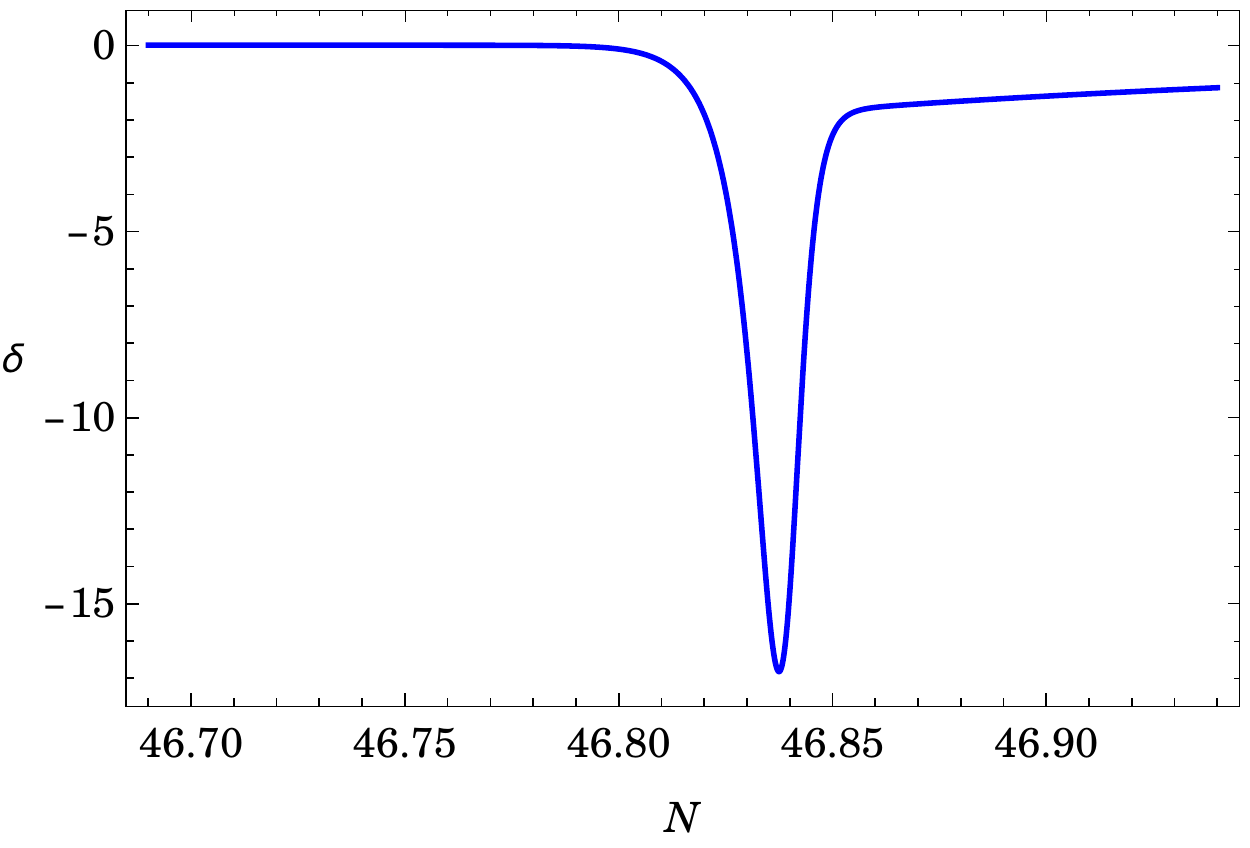}
 \qquad \includegraphics[scale=0.60]{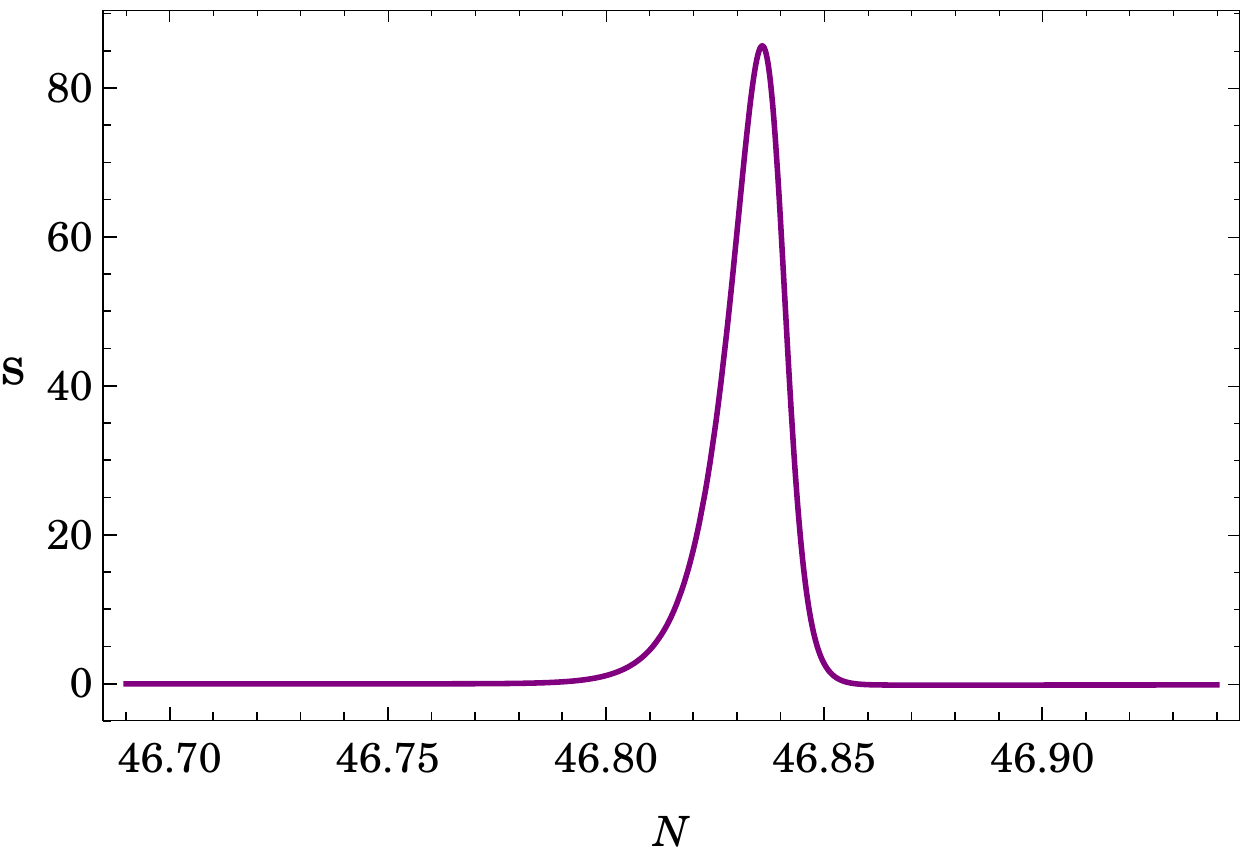}
\end{center}
\caption{Slow-roll parameters $\epsilon$,$\delta$ and $s$ as a function e-folds $N$ during DBI inflation for $\lambda=10^{10}$, $a=70$, $b=69$, $c=2000$, $d=6$.}\label{fig:DBI1}
\end{figure}

\begin{figure}[h!]
\begin{center}
\includegraphics[scale=0.68]{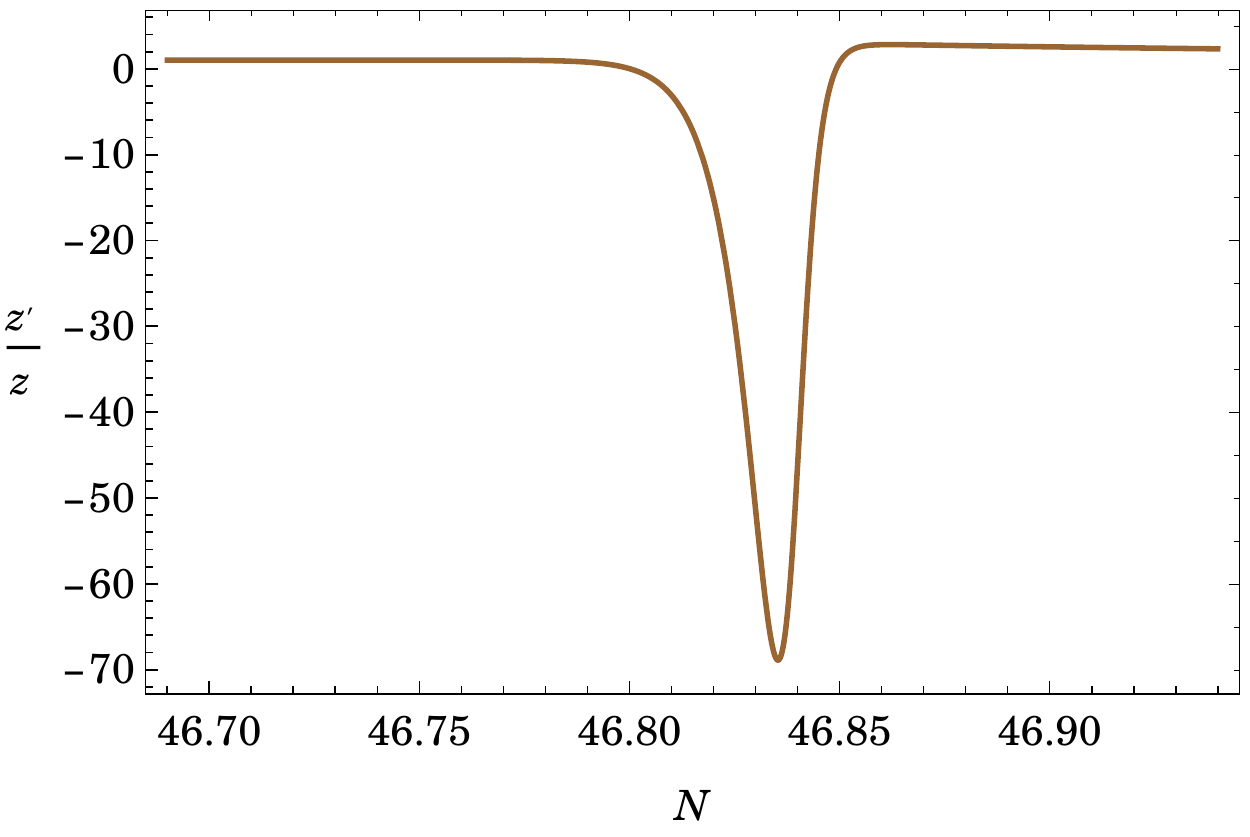} 
\end{center}
\caption{The behavior of $z'/z$ in units of the comoving horizon size $(aH)^{-1}$ around the step like feature in the warp factor. The parameter choices are the same as in Figure \ref{fig:DBI1}.}\label{fig:DBI3}
\end{figure}

As the   D3-brane moves down the throat, it  encounters a step downwards in the warp factor.  This step produces a sharp increase in the speed of sound, $c_s$, and thus a large, positive value for $s$.  So long as $\epsilon, \delta$ remain smaller than $s$, there will be a region where $z'/z <0$.  This gives rise to a large driving term in the equation governing the curvature perturbation \eqref{mez2}, and an expected growth in the superhorizon curvature perturbation.  Indeed the parameters in our model can easily be chosen to produce a transiently large and negative value for $z'/z$, and we give an illustrative example in Figures \ref{fig:DBI1}-\ref{fig:DBI3}.  

Note that this proposal is distinct from the one we made in Section \ref{Sec2}, in particular here there is no phase of ultra slow-roll, rather $\epsilon$ increases through the feature.  Also, compared to the models presented in Section \ref{Sec2}, $\delta<0$ and therefore $z'/z$ becomes negative due solely to the large positive value of $s$. Moreover $z'/z$ is negative for a much briefer time (in e-folds), but it can also be much more negative; thus here we have a much stronger driving force for a much briefer period of time.  Given the technical challenges in performing a complete numerical analysis for the cosmological perturbations in the present model with non-canonical kinetic terms including a sharp feature, we postpone this for future work.  However, the heuristic arguments presented here and in Sections \ref{SecHeu} and \ref{EPSPBH} suggest that an amplification of power in superhorizon modes can be achieved, and that this could provide a novel mechanism to produce interesting populations of PBHs.

\section{Conclusions}

In this paper, we present two well-motivated string theory scenarios which could give rise to a significant production of primordial black holes.  If observed, these primordial black holes would not only provide an explanation for the nature of some or all of the Dark Matter that dominates cosmic structures, but also provide another window into the early Universe, looking onto a period of the inflationary epoch unseen by the CMB.  Our interest in the current work has been in the production of PBHs around the mass scale $10^{17}-10^{18}~ {\rm g}$ ($10^{-16}-10^{-15}~M_{\odot}$), which can contribute a significant fraction of Dark Matter whilst evading current observational bounds.  For such a PBH population, the amplitude of primordial density perturbations at the scales of interest ($k \sim 10^{13}$ Mpc$^{-1}$) should be around $A_s \sim 10^{-2}$, whilst the CMB observations set the amplitude at $10^{-9}$ for scales around $k_* \sim 0.05$  Mpc$^{-1}$.  Therefore, PBH Dark Matter requires that the inflationary potential has some distinct behaviour between CMB and PBH scales, which ultimately should be explained within the underlying fundamental theory. 

A simple mechanism to enhance the amplitude of curvature perturbations during inflation within field theory models was identified by Leach et al \cite{Leach:2000yw, Leach:2001zf}.  The idea is that perturbations on super-horizon scales can undergo a large amplification when the slow-roll approximation does not apply.  Indeed, the equation governing the Fourier modes of the curvature perturbation takes the form of a damped harmonic oscillator:
\beq
\mathcal{R}''_k + 2\fr{z'}{z}\mathcal{R}'_k
 + k^2
\mathcal{R}_k = 0 \,,
\eeq
where $z \equiv a \dot{\phi}/H$.
  When the slow-roll approximation is a good one, around horizon crossing the dynamics are dominated by the friction term:
\beq
\fr{z'}{z}= aH \left(1+\epsilon-\delta \right)
\eeq
	and the solution is well-approximated by $\mathcal{R}_k$ constant, the growing adiabatic mode.  But if the slow-roll approximation temporarily breaks down -- in particular if $\delta > 1 + \epsilon$ -- then the friction term can momentarily change to a driving term, leading to an amplification of the modes that have recently left the horizon.  This may occur, for instance, when the potential has a region that is too steep to sustain slow-roll inflation and/or transitions to a region so flat that it supports an ultra slow-roll inflation.  We consider ways within string theory to realise this idea and produce in particular sufficient amplification at the right scales to produce PBH Dark Matter.

Our first example is given by string axion inflation, including significant but subleading non-perturbative corrections which add bumps to the leading monomial axion potential.  In earlier work, we showed that such effects can put axion inflation back into the favour of CMB observations, and here we note that they can moreover lead to an enhancement of power in the curvature perturbations at smaller scales.  Indeed, after CMB scales, the inflaton continues to roll down its bumpy potential through cliffs and plateaus towards its global minimum.  The sub-leading corrections can be such that successive plateaus are shallower and shallower until a local minimum, inflection point and maximum emerges.  As the inflaton traverses the local minimum climbing up the hill and passing through the inflection point, it undergoes a large deceleration, such that the second slow-roll parameter breaks the slow-roll condition, $\delta < 1$.  Subsequently, the system enters a phase of ``ultra slow-roll'' inflation, with first slow-roll parameter dramatically suppressed $\epsilon \sim e^{-(2\delta)N}$ while $\delta \gtrsim 3$.  It is also possible that the cliff preceding the local minimum is so steep that it can lead to a temporary interruption of inflation, where $\epsilon$ surpasses unity on the cliff before the rapid deceleration due to the inflection point.

We perform a numerical analysis of the linearized cosmological perturbations in our string-motivated model, using the MultiModeCode, and find the expected amplification of the curvature perturbation, in concordance with the heuristic arguments from \cite{Leach:2000yw, Leach:2001zf}.  This result for the primordial power spectrum can then be used to estimate the present day PBH abundance.  The fairly sharp peak in the curvature power spectrum leads to a fairly monochromatic PBH population.  The five parameters in our model can be adjusted such that the CMB observables are within 2$\sigma$ of the Planck data, and the mass and abundance of PBHs provide an order one fraction of the Dark Matter energy density. We also considered the implications of our scenario on the reheating epoch: for some -- but not all -- viable parameter choices an exotic post-inflationary equation of state is required.  
In addition to the population of light PBHs, potential signatures of such a scenario include a tensor-to-scalar ratio in the CMB of order $r \approx 10^{-3}-10^{-2}$ and a large negative running of the spectral index.

The second way to amplify the super-horizon curvature perturbations that we present is found within the class of DBI inflation.  Extending the Leach et al. argument to the case of DBI inflation, with a non-canonical kinetic term for the inflaton, reveals an additional contribution to the friction term in the mode equation:
\beq
\fr{z'}{z}= aH \left(1+\epsilon-\delta -s\right) \quad \text{where} \quad s = \frac{\dot c_s}{H c_s}\,.
\eeq
The speed of sound in DBI inflation is determined by the warp factor, $c_s = 1/\gamma = \sqrt{1 - h(\phi)\dot{\phi}^2}$.  Therefore, a steep downward step in the warp factor can lead to a large and positive $\dot{c_s}$, which transforms the friction term to a driving term and can thus potentially amplify the power in the super-horizon curvature mode.  For example, this feature in the warp factor may occur due to geometrical effects such as 
  a throat within a throat, sourced by two separated stacks of D-branes \cite{Franco:2005fd, Cascales:2005rj}.  Alternatively, a throat sourced by D-branes and/or fluxes would suffer an instability in the presence of $\overline{\text{D}}$-branes at its tip, due to perturbative brane-antibrane or non-perturbative flux-antibrane \cite{Kachru:2002gs} annihilation.  If the system underwent this instability towards less warping at some time during the inflationary trajectory, this would have interesting consequences on the inflationary perturbations.  

In the current paper, we model such phenomena by introducing a (smoothed) step-function into the warp factor.  By solving numerically the background field equations through this feature, we show that indeed it can lead to a driving term in the mode equation and propose that this could lead to significant enhancement of amplitude in the curvature power spectrum at corresponding late small scales.  Note that this would be distinct to previous scenarios in the literature, as there is no epoch of ultra-slow roll, rather $\epsilon$ increases through the feature.  Due to the technical challenges in solving for the cosmological perturbations in this model with a non-canonical kinetic term with sharp feature, we postpone a detailed numerical analysis for future work.  However, a rough quantatitive analysis of the behaviour of the curvature perturbation modes through the feature as in \cite{Leach:2000yw, Leach:2001zf}, supports the possibility that such a setup could provide a novel realisation of power enhancement in the primordial curvature perturbations and thus an interesting mechanism to produce PBHs.

Aside from a detailed numerical study of perturbations in our DBI model, there remain several interesting open questions. Our analysis on the dynamics in this work has been focused on the classical trajectory of the inflaton. However, in the region where inflaton experiences a strong deceleration, quantum fluctuations of the inflaton may dominate over extremely slow classical trajectory. In the slow-roll approximation, a recent analysis on this issue indeed show that quantum fluctuations can play a significant role in the estimates of PBH population \cite{Pattison:2017mbe}. Therefore, it would be interesting to study in some detail the role of these quantum fluctuations during the bumpy axion inflation we propose here, especially around the shallow minimum where slow-roll approximation breaks down.  Another issue under current discussion is the impact of non-Gaussianities in the density perturbations on the PBH abundances. 

 Specific to our string-inspired models,  it would be important to model more accurately the features in the warp factor that can arise in string compactifications. Although in principle the parameters $\{f,\Lambda_i, m\}$ of the axion model \eqref{pot} (or $\{c,d\}$ in the DBI model \eqref{WF}) could be tuned to adjust the position of the inflection point and produce a large population of PBH in the LIGO band ($M = 10-100 M_\odot$), we could not find a viable parameter space that simultaneously produces a large peak in the scalar power spectrum and agrees with the CMB observations. It could be interesting to pursue this possibility by focusing on other axionic potentials that might arise in string theory constructions which we will leave for future work. Moreover, an intriguing possibility is that multiple features in the potential or speed of sound might allow for multiple monochromatic populations of PBHs, thus helping to evade current observational bounds and provide sufficient abundancies to explain all of Dark Matter.  
Another interesting prospect is the possibility of an observable stochastic gravitational wave (GW) background associated with the scales related to the PBH formation, \eg sourced at second order by the large scalar perturbations (see also \cite{Cook:2011hg,Barnaby:2012xt,Ozsoy:2014sba,Namba:2015gja,Ozsoy:2017blg} for GW's sourced by amplified vector fields) which seed the PBHs \cite{Alabidi:2013wtp,Garcia-Bellido:2017aan}.

To summarise, we have shown that string inflationary models are rich enough to match CMB observations and produce PBH populations with masses  $\sim 10^{-16}-10^{-15}~M_{\odot}$ sufficient to explain Dark Matter.  Bounds on PBHs are consistently improving.  If such PBHs are observed in the future -- and note there are no known astrophysical mechanisms to produce black holes in this mass range -- they could not only explain the nature of Dark Matter but give invaluable information into the inflationary epoch.  Together with observables such as the tensor-to-scalar ratio, the running of the spectral indices and non-Gaussianities in the CMB, it could be possible to further narrow down the stringy mechanisms at play during the inflationary epoch.

\acknowledgments
O\"O, GT and IZ are partially supported by the STFC grant ST/P00055X/1. SP thanks to the Swansea University Physics Department for its hospitality during the initial stages of this project.

\addcontentsline{toc}{section}{References}

\bibliographystyle{utphys}

\bibliography{paper2}

\end{document}